\documentclass[journal]{IEEEtran}
\usepackage{cite}
\usepackage{amsmath,amssymb,amsfonts}
\usepackage{algorithmic}
\usepackage{graphicx}
\usepackage{algorithm,algorithmic}
\usepackage{hyperref}
\usepackage{textcomp}
\usepackage{amsmath, amssymb}
\usepackage{amsfonts}
\usepackage{xcolor}
\usepackage{siunitx}
\usepackage{standalone}
\usepackage{booktabs}
\usepackage{mdframed}
\usepackage{fancyvrb,multirow}
\usepackage{soul}
\usepackage{dsfont,mathabx}
\usepackage{array, booktabs}
\usepackage{booktabs}
\usepackage{makecell}
\usepackage{threeparttable}
\usepackage{bm}
\usepackage{multirow}
\usepackage{subfigure}
\usepackage{siunitx}
\usepackage{ulem}
\usepackage[english]{babel}
 
\newtheorem{theorem}{Theorem}

\newtheorem{lemma}{Lemma}

\newtheorem{definition}{Definition}
\newtheorem{remark}{Remark}

\def\BibTeX{{\rm B\kern-.05em{\sc i\kern-.025em b}\kern-.08em
    T\kern-.1667em\lower.7ex\hbox{E}\kern-.125emX}}
\usepackage{balance}

\begin{document}
\title{Curvature-Constrained Vector Field for Motion Planning of Nonholonomic Robots}
\author{Yike Qiao, \IEEEmembership{Student Member, IEEE}, Xiaodong He, \IEEEmembership{Member, IEEE}, An Zhuo, Zhiyong Sun, \IEEEmembership{Member, IEEE}, Weimin Bao and Zhongkui Li, \IEEEmembership{Senior Member, IEEE}

\thanks{Manuscript received March 25, 2025; revised August 21, 2025; accepted November 26, 2025. This work was supported in part by the National Natural Science Foundation of China under grants 62425301, U2241214, T2121002, 62503042, 62373008, and 62533017, in part by the Beijing Natural Science Foundation under grant 4254102.}
\thanks{Y. Qiao, A. Zhuo, Z. Sun and Z. Li are with the School of Advanced Manufacturing and Robotics, Peking University, Beijing 100871, China (e-mail: qiaoyike@stu.pku.edu.cn; zhuoan@stu.pku.edu.cn; zhiyong.sun@pku.edu.cn; zhongkli@pku.edu.cn)}
\thanks{X. He is with the School of Automation and Electrical Engineering, University of Science and Technology Beijing, Beijing 100083, China (e-mail: hxd@ustb.edu.cn)}
\thanks{W. Bao is with China Aerospace Science and Technology Corporation, Beijing 100048, China (e-mail: baoweimin@cashq.ac.cn)}
}

\maketitle
\begin{abstract}
Vector fields are advantageous in handling nonholonomic motion planning, as they provide the robot with reference orientation across the workspace. However, additionally incorporating curvature constraints presents challenges due to the interconnection between the design of the curvature-bounded vector field and the tracking controller under limited actuation.
In this paper, we present a novel framework to co-develop the vector field and the control laws, guiding the nonholonomic robot to the target configuration with curvature-bounded trajectory.
First, we formulate the problem by introducing the target positive limit set, which allows the robot to either converge to or pass through the target configuration, depending on its dynamics and the specific tasks. Next, we construct a curvature-constrained vector field (CVF) via blending and embedding elementary flows in the workspace. To track such CVF, the saturated control laws with a dynamic gain are proposed, under which the tracking error's magnitude decreases even when saturation occurs. Under the control laws, kinematically constrained nonholonomic robots are guaranteed to track the reference CVF and converge to the target positive limit set with bounded trajectory curvature.
Numerical simulations show that the proposed CVF method outperforms other vector-field-based algorithms. Experiments on Ackermann UGVs and semi-physical fixed-wing UAVs demonstrate that the method can be effectively implemented in real-world scenarios.


\end{abstract}


\section{Introduction}\label{sec:introduction}

\IEEEPARstart{M}{otion} planning, a fundamental problem in robotics, is critical for fulfilling various robotic tasks, such as autonomous driving, aerial surveillance, and underwater exploration. A key objective in motion planning is to provide feasible reference trajectories and control inputs to guide the robot to the target configuration.
To achieve feasibility and efficiency, motion planning algorithms should take into account the robot's kinematic constraints, among which the nonholonomic and curvature constraints are frequently encountered.
Nonholonomic constraints are typically imposed by the lateral velocity of robots being zero \cite{egeland1996feedback,li2013finite,he2024roto}, under which
the longitudinal linear velocity and angular velocity are considered as the control inputs in motion planning \cite{panagou2014motion,levin2019real,aiello2022fixed}.
In addition, as a result of the limited actuation of nonholonomic robots, such as unmanned ground vehicles (UGVs) with minimum turning radius \cite{balluchi1996optimal,chen2019shortest}, unmanned aerial vehicles (UAVs) with maximum lateral acceleration \cite{shima2002time,taub2013intercept}, and autonomous underwater vehicles (AUVs) with limited control moment \cite{du2016robust,li2019robust}, the trajectory curvature of these robots is subject to an upper bound, termed the curvature constraint.
The coexistence of the nonholonomic and curvature constraints poses significant challenges for motion planning, due to the tightly coupled position and orientation kinematics as well as the robots' limited actuation.

Existing motion planning methods addressing the above constraints fall into several categories, each of which has limitations that hinder its practical implementation.
Search-based methods are widely used, but they require post-processing and high computational cost to handle kinematic constraints \cite{murray1993nonholonomic,karaman2011sampling,song2018t}.
Optimization-based methods can effectively manage various constraints, but often suffer from issues related to local minima and substantial computational burdens \cite{Hussein2008Optimal,duan2014planning,xu2021autonomous}.
Although the Dubins curve-based method is a well-studied algorithm for generating curvature-bounded trajectories for nonholonomic robots, it does not incorporate control input design for path following \cite{dubins1957curves,chen2020dubins,chen2023elongation}.
Moreover, the common drawback of the aforementioned methods is that they operate in an open-loop manner, necessitating replanning when deviations occur.

Unlike the preceding open-loop methods, the vector field (VF)-based approaches specify the desired orientation based on the nonholonomic robot's position by assigning a vector to each point in the workspace.
One classic way to generate VFs is by calculating the gradient of potential functions, known as the artificial potential field (APF) \cite{tanner2003nonholonomic,pathak2005integrated,masoud2009harmonic}. However, this method is prone to local minima, which possibly leads to deadlock.
To address this problem, Lindemann \textit{et al.} utilize cell decomposition approach to design a non-gradient VF for each polygonal cell and achieve convergence to the target position \cite{lindemann2006real,lindemann2007smooth,lindemann2009simple}.
With dipole-like VF for guidance, Panagou \textit{et al.} further derive the tracking control laws for the VF to stabilize a nonholonomic robot \cite{panagou2010dipole}, and coordinate the motion of multiple unicycles \cite{Panagou2017Distributed}.
He \textit{et al.} \cite{he2025novel} generalize dipole-like VF to 3D motion planning, guiding multiple nonholonomic robots to their target positions with a specific heading direction except for a singular set of zero measures.
To address the issues that arise from the singularity of VFs, singularity-free VFs and control laws for fixed-wing UAVs are proposed to achieve path following and motion coordination in \cite{yao2021singularity,yao2023guiding}.

Although the above VFs naturally address the nonholonomic constraint, few existing works incorporate the curvature constraint into the VF-based motion planning. The challenges in addressing both the nonholonomic and curvature constraints lie in at least two aspects.
On the one hand, designing a curvature-constrained vector field (CVF) is far from trivial, particularly for the simultaneous position and orientation planning problem as discussed in \cite{he2024simultaneous}.

When the robot's orientation aligns with the VF, the robot's trajectory is the same as the VF's integral curves by moving along the VF.
To ensure the boundedness of the curvature of the trajectory and the continuity of the control inputs, the curvature of the VF should be not only bounded but also continuous\cite{scheuer1997continuous}.
Furthermore, navigating towards the target configuration requires the VF to coincide with the target orientation while approaching the target position, which becomes challenging since the adjustment of the orientation of the VF is limited by the curvature constraint.
For example, the VFs proposed in \cite{lau2015fluid} would converge to the desired position but do not guarantee alignment with a specified orientation, further highlighting the difficulty of VF design in simultaneous position and orientation planning.

On the other hand, the design of tracking control laws for the CVF becomes significantly more difficult under the nonholonomic and curvature constraints. 
In most cases, a nonholonomic robot's orientation differs from the CVF, leading to the deviation of its trajectory from the CVF's integral curve. 
Moreover, the curvature constraint imposes an upper bound proportional to the linear velocity on the angular velocity, restricting the robot from aligning its orientation with the CVF and hence exacerbating the deviation.
Take \cite{pothen2017curvature,shivam2021curvature}, for example, although their VFs have prescribed curvature limits, the control laws designed to track these VFs do not account for the limited actuation mentioned above. Hence, the actual trajectories may violate the curvature constraint when deviations occur.
To ensure the existence of feasible
control laws that can track the VF under curvature constraint, the co-design of the VF and control laws is crucial. 
Due to the aforementioned challenges, developing a feedback motion planning method to guide a nonholonomic robot with curvature constraints to the target configuration remains unsolved.

In this paper, we formulate the simultaneous position and orientation planning as guiding the curvature-constrained nonholonomic robot to the target positive limit set, a positive limit set containing the target configuration. To achieve this objective, we aim to find the reference trajectories and control inputs that satisfy the curvature constraint while converging to the target positive limit set.
This formulation is more general and applies to real-world scenarios, including Ackermann UGV converging to a configuration, and fixed-wing UAV passing through a configuration with nonzero velocity. In order to address the difficulties mentioned above, we propose a unified VF-based motion planning framework, under which the VF provides the curvature-bounded reference trajectories and the control laws ensure that the nonholonomic robot tracks the proposed VF under the curvature constraint.

The theoretical contributions of this paper are summarized as follows:
\begin{itemize}
    \item We construct a curvature-constrained vector field (CVF) with a stable limit cycle as the target positive limit set. A novel VF is proposed by partitioning the workspace and embedding the blended source, sink and vortex flows into different regions with conveniently tunable parameters.
    Moreover, we derive the sufficient condition ensuring that the CVF's integral curves have a prescribed curvature bound and serve as the reference trajectories converging to the limit cycle except at an isolated singular point. 
    Compared with the dipole-like VF in \cite{panagou2010dipole,he2025novel} with singular target position and the curvature-constrained VF in \cite{shivam2021curvature,pothen2017curvature}, this construction strategy reduces the effect of singularity during trajectory convergence, while allowing a flexible selection of CVF parameters, facilitating the co-design of CVF and control laws.
    \item We design novel control laws comprising the linear velocity with specified bounds adaptive to the planning objective and the saturated angular velocity, which ensures that the actual trajectory curvature is also bounded when tracking the CVF. To guarantee that the closed-loop system tracks the CVF under saturation of angular velocity, we propose a state-dependent dynamic gain that responds to changes in the CVF. In this way, the orientation error between the robot and CVF is monotonically stabilized even when saturation occurs.
    \item We unify the co-design of CVF and control laws under curvature constraint by establishing an explicit condition on CVF's parameters, to guarantee the existence of control inputs that track the CVF.
    This condition reveals the intrinsic coupling between the CVF and control laws under curvature constraint, and limits the orientation changing rate of CVF by allocating sufficient space in the blending regions of flow fields for a gradual transition. 
    Based on above results, we prove that the closed-loop system almost globally converges to the target positive limit set with bounded trajectory curvature.
  
\end{itemize}

To validate the theoretical results and facilitate reproducibility, we conduct and open source the implementations, including the numerical simulations and hardware experiments\footnote{Implementation details and source code are available at \url{https://github.com/Y1kee/CVF-for-Nonholonomic-Motion-Planning}}, 
as outlined below. 
\begin{itemize}
  \item We carry out numerical simulations demonstrating supportive phenomena for the theoretical propositions and Monte Carlo experiments comparing our method to other VF-based algorithms in \cite{Panagou2017Distributed,he2024simultaneous,pothen2017curvature,yao2021singularity}. The presented approach outperforms the existing methods by achieving a 100\% success rate in ensuring convergence under curvature constraints.
  \item We implement the proposed planning algorithm on the Ackermann UGV featuring a minimal turning radius. The experiment results show that they successfully converge to different destinations from various initial conditions, demonstrating that the proposed feedback motion planning algorithm is robust against the dynamics uncertainties.
  \item We extend the proposed method to 3D fixed-wing UAVs with coordinated turn constraint by deriving the attitude and thrust control laws. In the hardware-in-the-loop (HIL) experiments with target configuration switching during the flight, the UAV reaches its destination under the extended control laws. The HIL results highlight that our algorithm given by closed-form expressions enables real-time motion planning despite limited onboard computational resources.
\end{itemize}

The paper proceeds as follows.
Section~\ref{sec:preliminaries} presents the nonholonomic kinematics with curvature constraint and formulates the motion planning problem. 
Then, Section~\ref{sec:VF} shows how to build the CVF and demonstrates its desired properties of convergence and bounded curvature. 
To track the proposed CVF, saturated control laws with dynamic gains are introduced in Section~\ref{sec:control}, where almost global convergence under curvature constraint is established. 
Section~\ref{sec:numerical simulation} conducts numerical simulations to validate the theoretical results and compares our algorithm with other existing VF-based motion planners. 
In Section~\ref{sec:Experiments}, the proposed method is further evaluated through experiments using an Ackermann UGV and a semi-physical fixed-wing UAV with curvature-bounded trajectories.
Finally, Section~\ref{sec:conclusion} concludes this paper.

\section{Kinematic Model and Problem Statement}\label{sec:preliminaries}

\subsection{Modeling of Kinematics and Constraints}
In this paper, we let $\bm{g}=(\bm{p},\theta)$ denote the configuration of a 2D mobile robot, where $\bm{p}=[x,y]^\intercal\in\mathbb{R}^2$ is the position and $\theta\in[0,2\pi)$ is the orientation.
For the considered configuration, the nonholonomic constraint is imposed by confining the translational motion along the robot's heading direction, i.e., the $x$-axis of the body-fixed frame. In this case, the most commonly used nonholonomic kinematics are expressed as
\begin{equation}\label{nonholonomic kinematics}
  \begin{aligned}
    &\dot{x}=v_x\cos{\theta},\\
      &\dot{y}=v_x\sin{\theta},\\
      &\dot{\theta}=\omega,
  \end{aligned}
\end{equation}
where the control inputs consist of the linear velocity $v_x$ and the angular velocity $\omega$.

Due to the limitation on the steering mechanism of nonholonomic robots, the trajectory curvature is subject to an upper bound, known as the curvature constraint.
For kinematics depicted by \eqref{nonholonomic kinematics}, the curvature constraint is written as
\begin{equation}\label{curvature_constraint}
  \kappa = \left|\frac{ \omega }{v_x}\right| \le \bar{\kappa},
\end{equation}
where $\kappa$ is the curvature of robot's actual trajectory and $\bar{\kappa}$ is the maximum curvature. Additionally, the minimum turning radius of the nonholonomic robot is denoted by
\[\rho=1/\bar{\kappa}.\]

The above kinematic model captures the essential constraints for us to conduct rigorous analysis of the algorithm to be proposed, and serves as a foundation extendable to more complex robotic systems. The experiment results demonstrate that the proposed method is applicable to UGVs with Ackermann steering geometry and fixed-wing UAVs with banking turn control featuring cruising speed constraints.

\subsection{Problem Formulation}\label{ssec:formulation}
In real-world scenarios, mobile robots are usually required to reach the target position with predefined orientation to accomplish specific tasks.
For robots with positive lower bound on linear velocity due to task requirements or dynamics constraints, the convergence to a specific configuration is not always achievable.
For a more general problem formulation, we employ the concept of positive limit set in \cite[Ch. 4.2]{khalil2002nonlinear} and define the target positive limit set as follows.

\begin{definition} \label{positive limit set}
The positive limit set of the trajectory $\phi(t;\bm{g}_0)$ with initial configuration $\bm{g}_0$ is defined as
\[\mathcal{L}^+(\bm{g}_0) = \left\{ \bm{g} \mid \exists \{t_k\}_{k=1}^{\infty}\to \infty, \phi(t_k;\bm{g}_0)=\bm{g} \text{ as } k \to \infty. \right\}\]
\end{definition}

\begin{definition} \label{target positive limit set}
  The target positive limit set $\mathcal{L}_d^+(\bm{g}_0)$ is the positive limit set that contains the target configuration $\bm{g}_d$, i.e., $\bm{g}_d \in \mathcal{L}_d^+(\bm{g}_0)$, implying that
  \[\exists \{t_k\}_{k=1}^{\infty}\to \infty, \phi(t_k;\bm{g}_0)=\bm{g}_d \text{ as } k \to \infty. \]
\end{definition}


According to the above definitions, a controlled robot could approach the target configuration by converging to the target positive limit set.
To achieve this objective, the motion planner aims to derive curvature-bounded reference trajectories that converge to the target positive limit set, along with control inputs that track these reference trajectories. In this way, the nonholonomic system \eqref{nonholonomic kinematics} subject to the curvature constraint \eqref{curvature_constraint} converges to the target positive limit set, i.e.,
\begin{equation}\label{planning obj}
    \lim_{t\to\infty} d(\bm{g}(t),\mathcal{L}_d^+(\bm{g}_0))=0,
\end{equation}
where $d(\bm{g}_1,\mathcal{G})=\inf_{\bm{g}_2 \in \mathcal{G}}\sqrt{\|\bm{p}_1-\bm{p}_2\|^2+(\theta_1-\theta_2)^2}$ denotes the distance from $\bm{g}_1$ to the set $\mathcal{G}$.

\begin{remark}
  For nonholonomic robots like wheeled UGVs, they are able to converge to certain configuration with velocity approaching zero.
  In this case, the problem formulation is reduced to $\mathcal{L}^+(\bm{g}_0)=\{\bm{g}_d\}$ and $\lim_{t\to\infty}\bm{g}(t)=\bm{g}_d$.
  Besides, there exist dynamics constraints and task requirements that require non-zero velocity at the target configuration, such as fixed-wing UAVs and glider AUVs. 
  It makes sense to guide these robots to the target positive limit set that contains the target configuration, i.e., $\left\{\bm{g}_d\right\} \subset \mathcal{L}_d^+(\bm{g}_0)$, and hence they reach the target configuration repeatedly, i.e., $\exists \{t_k\}_{k=1}^{\infty}\to \infty$ such that $\bm{g}(t_k)=\bm{g}_d$ as $k \to \infty.$
\end{remark}

\subsection{Proposed Algorithm Framework}\label{spaf}
\begin{figure}[hptb!]
  \centering
  \includegraphics[width=0.8\linewidth]{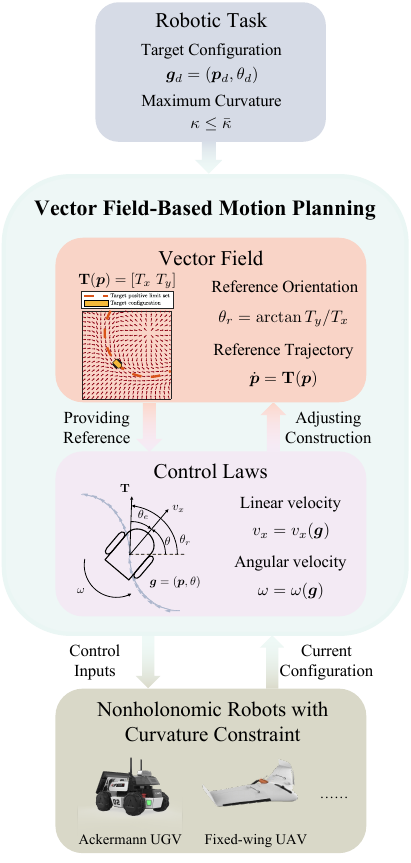}
  \caption{Diagram of the co-design methodology of vector field and control laws for motion planning presented in this paper.}
  \label{fig:flowchart}
\end{figure}

To address the motion planning problem formulated above, the algorithm to be proposed in this paper employs a feedback structure as illustrated in Figure~\ref{fig:flowchart}.
In this framework, the reference trajectory and reference orientation are provided by the vector field presented in Section~\ref{sec:VF} and the control laws to track the vector field are derived in Section~\ref{sec:control}.
Due to the co-design strategy, the vector field adjusts its parameters to ensure the boundedness of control inputs and the control laws updates the feedback gain in response to the change of vector field. 
In this manner, the robot tracks the vector field and converges to the target positive limit set with prescribed curvature bound.

\section{Curvature-Constrained Vector Field}\label{sec:VF}

In this section, we will construct the curvature-constrained vector field (CVF), where a stable limit cycle that contains the target configuration is established as the target positive limit set. To provide the robot with curvature-bounded reference trajectory, we further derive the condition to ensure that the integral curves of CVF have a prescribed upper bound on the curvature while converging to the limit set.

Before moving forward, we present several concepts about the vector field (VF) \cite[Chs.~8, 9]{lee2012introduction}.

\begin{definition}\label{def_VF}
  The vector field is a map $\mathbf{F}:\mathbb{R}^n\to\mathbb{R}^n$ and a vector field is of class $C^k$ if each component of the vector field has continuous derivatives up to order $k$.
\end{definition}

\begin{definition}
  A point $\bm{p}$ is a singular point for the vector field $\mathbf{F}$, if $\mathbf{F}(\bm{p})=\bm{0}$.
\end{definition}

\begin{definition}\label{def_IC}
  The integral curve $\bm{\zeta}(t)$ of a vector field $\mathbf{F}$ defined on $t\ge0$ is a map $\bm{\zeta}\colon \mathbb{R}_{\ge0} \to \mathbb{R}^n$ that satisfies
  \[\frac{\mathrm{d}\bm{\zeta}}{\mathrm{d}t}=\mathbf{F}(\bm{\zeta}),\ \bm{\zeta}(0)=\bm{\zeta}_0.\]
\end{definition}

For simplicity, we consider the polar coordinates $(r,\varphi)$ transformed by $r=\sqrt{x^2+y^2}$ and $\varphi=\arctan{(y/x)}$. Denote the basis of the Cartesian and polar coordinates as $\left\{\bm{e}_x,\bm{e}_y\right\}$ and $\left\{\bm{e}_r,\bm{e}_\varphi\right\}$, respectively. Then the components of the vector field represented in the polar coordinates are defined as $F_r=\left<\bm{e}_r,\mathbf{F}\right>$ and $F_\varphi=\left<\bm{e}_\varphi,\mathbf{F}\right>$, where $\left<\cdot,\cdot\right>$ is the inner product. 

Since the VF designates the desired velocity direction of the robot by assigning a vector to each position in the workspace, the integral curves could be regarded as the reference trajectories for the robot. By aligning the orientation with the VF, the nonholonomic robots' actual trajectories converge to the integral curves.
Therefore, to navigate a curvature-constrained nonholonomic robot, the integral curves should not only converge to the target positive limit set but also have prescribed bound on the curvature.
In the following lemma, the computation of curvature of integral curves is shown.

\begin{lemma}\label{lemma_curvature}
  Consider a VF $\mathbf{F}:\mathbb{R}^2\to\mathbb{R}^2$ given in the polar coordinates $(r,\varphi)$. The curvature of its integral curves at the nonsingular point $\boldsymbol{p}$ is given by
    \begin{equation}\label{VF_k}
        \kappa_\mathbf{F}\left(\boldsymbol{p}\right) = \frac{\lvert \left<\mathbf{F} ,\boldsymbol{K}_\mathbf{F} \mathbf{F}\right> \rvert}{\left \| \mathbf{F}\left(\boldsymbol{p}\right) \right \|^3},
    \end{equation}
    where $\boldsymbol{K}_\mathbf{F}$ is defined by
    \begin{equation}
      \boldsymbol{K}_\mathbf{F}=\left[ \begin{matrix}
        \frac{\partial{F_\varphi}}{\partial{r}} + \frac{F_\varphi}{r} & -\frac{\partial{F_r}}{\partial{r}} \\
         \frac{1}{r}\frac{\partial{F_\varphi}}{\partial{\varphi}} & -\frac{1}{r}\frac{\partial{F_r}}{\partial{\varphi}} + \frac{F_\varphi}{r}
      \end{matrix}\right].
    \end{equation}
\end{lemma}
\begin{IEEEproof}
  See Appendix~\ref{appendix1}.
\end{IEEEproof}

\begin{figure}[htbp]
  \centering
  \subfigure[Vortex $\mathbf{F}_\text{v}$]{\includegraphics[height=0.38\linewidth]{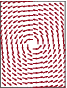} \label{fig:vortex}}
  \hfill
  \subfigure[Source $\mathbf{F}_\text{out}$]{\includegraphics[height=0.38\linewidth]{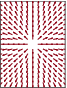}\label{fig:source}}
  \hfill
  \subfigure[Sink $\mathbf{F}_\text{in}$]{\includegraphics[height=0.38\linewidth]{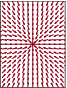} \label{fig:sink}}
  \caption{Three types of elementary flow with singular point at the origin.}
  \label{fig:planar flow fields}
\end{figure}

To construct a desired CVF with the aforementioned properties, we utilize the elementary flows of vortex, source and sink, which can constitute a wide variety of VFs by superposition\cite[Ch. 8]{White2016Fluid}.
Specifically, the normalized elementary flows are given, respectively, by the following equations:
\begin{equation}\label{vortex}
  \mathbf{F}_\text{v}=\left\{\begin{matrix}
  &\left[F_{\text{v},r},F_{\text{v},\varphi}\right]^\intercal=\left[0,1\right]^\intercal, & \bm{p}\neq\bm{0},\\
  &\bm{0}, &\bm{p}=\bm{0}.
  \end{matrix}\right.
\end{equation}
\begin{equation}
  \mathbf{F}_\text{out}=\left\{\begin{matrix}
      &\left[F_{\text{out},r},F_{\text{out},\varphi}\right]^\intercal=\left[1,0\right]^\intercal,
      & \bm{p}\neq\bm{0},\\

      &\bm{0}.
      &\bm{p}=\bm{0}.
  \end{matrix}\right. \label{source}
\end{equation}
\begin{equation}
    \mathbf{F}_\text{in}=\left\{\begin{matrix}
        &\left[F_{\text{in},r},F_{\text{in},\varphi}\right]^\intercal=\left[-1,0\right]^\intercal,
        & \bm{p}\neq\bm{0},\\

        &\bm{0}.
        &\bm{p}=\bm{0}.
    \end{matrix}\right.\label{sink}
\end{equation}
Figure~\ref{fig:planar flow fields} shows the above elementary flows, where the integral curves of the vortex are concentric circular arcs around the singular point, while the integral curves of source and sink are straight lines diverging from and converging to the singular point, respectively. 
The geometric forms of these integral curves match the motion pattern of nonholonomic robots, since the trajectory of such systems can generally be described as compositions of circular arcs and straight-line segments.

Drawing intuition from the phase portrait of a nonlinear system featuring a stable limit cycle, we construct a desirable VF with the following transition of the elementary flows as the distance from the singular point increases: from the sink near the singularity, to the vortex around the limit cycle, and finally to the source further out.
To achieve this transition, we partition the workspace with three concentric circles with radii $0<r_1<r_2<r_3$:
\begin{equation}\label{decomposition}
    \begin{aligned}
      &\mathcal{A}_1=\{\bm{p} \mid 0\le r < r_1\},\\
      &\mathcal{A}_2=\{\bm{p} \mid r_1 \le r < r_2\},\\
      &\mathcal{A}_3=\{\bm{p} \mid r_2 \le r < r_3\},\\
      &\mathcal{A}_4=\{\bm{p} \mid r \ge r_3 \},
    \end{aligned}
\end{equation}
where $\mathcal{O}_2=\left\{r=r_2\right\}$ is targeted as the stable limit cycle with the singular point at the origin.
Given the above decomposition, we embed the source and sink in the regions $\mathcal{A}_1$ and $\mathcal{A}_4$, and then blend the vortex with source and sink in the regions $\mathcal{A}_2$ and $\mathcal{A}_3$, respectively.
Consider a blending function $\lambda(s;s_l,s_u)$ that monotonically decreases from 1 to 0 on its domain $[s_l,s_u]$ and denote $\tilde{\lambda}(s;s_l,s_u)=1-\lambda(s;s_l,s_u)$. The blended VF can be given by 
\begin{equation}\label{the_VF}
  \mathbf{F}(\bm{p}) = 
  \begin{cases}
    \mathbf{F}_\text{out}, & \bm{p}\in\mathcal{A}_1, \\
    \mathbf{F}_\text{out}^{\text{v}}, & \bm{p}\in\mathcal{A}_2, \\
    \mathbf{F}_\text{v}^\text{in}, & \bm{p}\in\mathcal{A}_3, \\
    \mathbf{F}_\text{in}, & \bm{p}\in\mathcal{A}_4,
  \end{cases}
\end{equation}
where the transition between elementary flows is specified by
\begin{equation}
  \mathbf{F}_\text{out}^\text{v}=\lambda(r;r_1,r_2)\mathbf{F}_\text{out}+\tilde{\lambda}(r;r_1,r_2)\mathbf{F}_\text{v},\
  \bm{p}\in\mathcal{A}_2, \label{out,v}
\end{equation}
\begin{equation}
  \mathbf{F}_\text{v}^\text{in}=\lambda(r;r_2,r_3)\mathbf{F}_\text{v}+\tilde{\lambda}(r;r_2,r_3)\mathbf{F}_\text{in},\
  \bm{p}\in\mathcal{A}_3. \label{v,in}
\end{equation}

According to Definition \ref{def_VF} and Lemma \ref{lemma_curvature}, the above blended VF has continuous curvature, which is a requirement for the continuity of the tracking control inputs, when the VF is of class $C^1$ at least. 
By calculating the partial derivatives of \eqref{the_VF}, we can find that the blended VF is of the same class $C^k$ as the blending function within $\mathcal{A}_{i=2,3}$.
Moreover, the blended VF has continuous partial derivatives on the boundaries between $\mathcal{A}_{i=1,2,3,4}$ only when the blending function has zero derivatives at these boundaries.
Therefore, the necessary and sufficient condition for the blended VF to have continuous curvature, is that the blending function $\lambda(r;r_l,r_u)$ is of class $C^{k\ge 1}$ and $\frac{\partial \lambda}{\partial r}\mid _{r=r_l} = \frac{\partial \lambda}{\partial r}\mid _{r=r_u} = 0$.

In this paper, we employ the blending function given by
\begin{equation}\label{blend}
  \lambda(s;s_l,s_u)=2\left(\frac{s-s_l}{s_u-s_l}\right)^3 - 3\left(\frac{s-s_l}{s_u-s_l}\right)^2+1,
\end{equation}
which smoothly decreases from 1 to 0 on the interval $[s_l,s_u]$ and has zero derivatives on the boundaries. 

\begin{figure}
  \centering
  \includegraphics[width=0.7\linewidth]{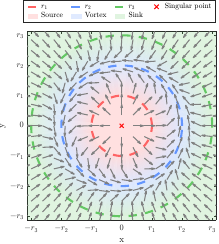}
  \caption{The blended vector field $\mathbf{F}$, where the domains of source, vortex and sink are shaded in red, blue and green. 
  Within the transition region between red and green dotted circles, the blended vector field evolves from source to vortex and then to sink as the distance from the singular point increases.}
  \label{fig:blended VF}
\end{figure}

It is worth noting that the derivative of this blending function is related to the width of domain. 
As a result, the transition behavior between different elementary flows in the blended VF can be easily adjusted by modifying the region radii $r_i$ as shown in Figure~\ref{fig:blended VF}.
Based on appropriate selection of $r_i$, we ensure that the integral curve of the blended VF has a prescribed curvature bound $\bar{\kappa}$ while converging to the limit cycle. This is formally stated in the following theorem.

\begin{theorem}\label{thm F}
The blended vector field defined in \eqref{the_VF} satisfies the following properties:
\begin{enumerate}
  \item $\mathbf{F}$ has a unique and isolated singular point at $\bm{p}=\bm{0}$ and $\mathbf{F}$ is of class $C^1$ over $\mathcal{D}=\mathbb{R}^2\setminus \bm{0}$;
  \item $\mathcal{O}_2$ is an almost-globally stable limit cycle for the integral curves with dynamics $\dot{\bm{\zeta}}=\mathbf{F}(\bm{\zeta})$;
  \item The integral curves of $\mathbf{F}$ have continuous curvature bounded by $\bar{\kappa}$, if 
\end{enumerate}
\begin{subequations}\label{curvature condition}
  \begin{align}
    r_{i+1} - r_i \ge 3\rho, \quad i=1,2. \label{curvature condition 2}\\
    r_i \ge r_{i+1}/2, \quad i=1,2. \label{curvature condition 1}
  \end{align}
\end{subequations}
\end{theorem}

\begin{IEEEproof}
    See Appendix~\ref{appendix_F}.
\end{IEEEproof}

\begin{remark}\label{remark_radius}
  The physical meaning of the condition \eqref{curvature condition} is explained as follows.
  In $\mathcal{A}_2$ and $\mathcal{A}_3$, the vector field gradually evolves from the source and sink to the vortex as it approaches the limit cycle.
  As the regions $\mathcal{A}_2$ and $\mathcal{A}_3$ expand, the transition becomes smoother and the curvature in the blended region decreases.
  Condition \eqref{curvature condition 2} allocates enough space for the blended vector field to steer its orientation with bounded curvature when approaching the limit cycle.
  Moreover, the domain of vortex should be expelled away from the singular point by increasing the region radii $r_i$, since the curvature of the vortex becomes unbounded around the singular point.
  The condition \eqref{curvature condition 1} imposes a lower bound on the region radii such that the domain of vortex in the blended VF, i.e., $\mathcal{A}_2\bigcup\mathcal{A}_3$, is far enough from the singular point to avoid violation of the curvature constraint.
\end{remark}

Since the stable limit cycle $\mathcal{O}_2$ is circular, we can always find a point $\bm{p}_{\theta_d}\in \mathcal{O}_2$ where the tangent vector aligns with the target orientation $\theta_d$, i.e., $\mathbf{F}(\bm{p}_{\theta_d})=[\cos\theta_d,\sin\theta_d]^\intercal$. By translating the $\bm{p}_{\theta_d}$ to the target position $\bm{p}_d$, we establish the limit cycle as the target positive limit set and present the CVF as below:
\begin{equation}\label{CVF}
  \mathbf{T}(\bm{p}) = \left\{\begin{matrix}
  & \frac{\mathbf{F}(\bm{p}-\bm{p}_\delta)}{\|\mathbf{F}(\bm{p}-\bm{p}_\delta)\|},
  & \bm{p}\neq\bm{p}_\delta,\\

  & \bm{0},
  & \bm{p}=\bm{p}_\delta,
  \end{matrix}\right.
\end{equation}
where $\bm{p}_\delta=\bm{p}_d-\bm{p}_{\theta_d}$ is the singular point and $\bm{p}_{\theta_d}=r_2 [\cos(\theta_d-\pi/2),\sin(\theta_d-\pi/2)]^\intercal$.

The properties of the CVF are summarized in the following theorem.

\begin{theorem}\label{thm T}
  The curvature-constrained vector field given by \eqref{CVF} has the following properties:
    \begin{subequations}
    \begin{align}
      &\kappa_{\mathbf{T}}(\bm{p}) \in C^0,\ \kappa_\mathbf{T}(\bm{p}) \le \bar{\kappa},\ \forall \bm{p} \neq \bm{p}_\delta, \label{CVF curvature} \\
      &\bm{p}_d \in \mathcal{L}^+_\mathbf{T}(\bm{\zeta}_0),\ \forall \bm{\zeta}_0 \neq \bm{p}_\delta, \label{CVF position} \\
      &\mathbf{T}(\bm{p}_d)=\left[\cos{\theta_d},\sin{\theta_d}\right]^\intercal, \label{CVF orientation}
    \end{align}
  \end{subequations}
  where $\mathcal{L}^+_\mathbf{T}(\bm{\zeta}_0)$ is the positive limit set of the integral curves.
\end{theorem}

\begin{IEEEproof}
  See Appendix~\ref{appendix_T}.
\end{IEEEproof}

\begin{figure}
  \centering
  \includegraphics[width=0.7\linewidth]{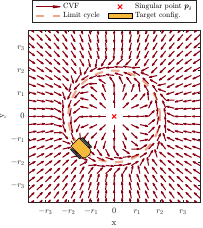}
  \caption{The proposed CVF $\mathbf{T}$, where the singular point is the origin.}
  \label{fig:T}
\end{figure}

The proposed CVF is illustrated by Figure~\ref{fig:T}. In the above theorem, the property~\eqref{CVF curvature} means that the curvature of integral curves of the CVF is continuous and bounded by $\bar{\kappa}$ for any non-singular position. 
The property~\eqref{CVF position} indicates that the integral curves of the CVF converge to a stable limit cycle containing the target position $\bm{p}_d$ almost globally. Moreover, the property~\eqref{CVF orientation} ensures that the orientation of the CVF aligns with the target orientation $\theta_d$ at $\bm{p}_d$.
By moving along the integral curves of the CVF, the nonholonomic robot approaches the target configuration while converging to the target positive limit set with continuous trajectory curvature bounded by $\bar{\kappa}$.


\begin{remark}
Compared to existing VF-based motion planners with curvature constraints, the proposed CVF offers several key advantages. In \cite{lau2015fluid}, the VF only converges to the target position and depends on empirical parameter tuning. In contrast, our CVF guarantees that curvature-bounded integral curves simultaneously plan both position and orientation, as specified by the explicit condition in \eqref{curvature condition}.
While \cite{qiao2024motion} also addresses position and orientation planning, it places the singularity of the curvature-constrained VF at the target position. Unlike this approach, our CVF establishes a stable limit cycle as the target positive limit set, keeping the target configuration away from the singular point and avoiding singularity during convergence.
Although \cite{pothen2017curvature} utilizes a stable limit cycle, the limited flexibility in its VF design makes it challenging to derive control inputs that both track the VF and maintain the curvature constraint. In contrast, our CVF offers more flexibility in parameter tuning, facilitating the co-design of the CVF and control laws in the subsequent section.
\end{remark}

\section{Saturated Control Laws with Dynamic Gain}\label{sec:control}

The CVF presented in Theorem~\ref{thm T} provides reference trajectories with desired curvature bound, converging to the target positive limit set. As pointed out in Section \ref{spaf}, we will in this section find feasible control inputs of the nonholonomic robot, including the linear velocity $v_x$ and angular velocity $\omega$, to track the CVF under curvature constraint.

In accordance with the motion planning objective, we firstly specify the control law with user-defined bounds as follows:
\begin{equation}\label{vx}
  v_x=v_{\text{min}}+k_v \tanh{\left( \frac{\|\bm{p}-\bm{p}_d\|}{c_{\bm{p}}} + \frac{\lvert \theta_e \rvert}{c_\theta} \right)},
\end{equation}
where $v_{\text{min}}$ is the lower bound, $c_{\bm{p}}$, $c_\theta$ are positive scalars, and the gain is given by $k_v=v_{\text{max}}-v_{\text{min}}$ such that the linear velocity is upper bounded by $v_{\text{max}}$. Moreover, $\theta_e$ is the orientation error between the robot and CVF defined by
\begin{equation} \label{theta_e}
  \theta_e=\theta-\theta_r,
\end{equation}
where the reference orientation $\theta_r$ of CVF is specified as
\begin{equation}\label{theta_ref}
  \theta_r(\bm{p})=\arctan{\frac{\left<\mathbf{T}(\bm{p}),\bm{e}_y\right>}{\left<\mathbf{T}(\bm{p}),\bm{e}_x\right>}},\ \bm{p}\neq\bm{p}_\delta.
\end{equation}
Since the $\bm{p}_\delta$ is the singular point of the CVF, in the following we adopt the polar coordinates centered at $\bm{p}_\delta$:
\[
\begin{aligned}
    r_\delta &= \|\bm{p}-\bm{p}_\delta\|, \\
    \varphi_\delta &= \arctan{\frac{\left<\bm{p}-\bm{p}_\delta,\bm{e}_y\right>}{\left<\bm{p}-\bm{p}_\delta,\bm{e}_x\right>}}.
\end{aligned}
\]
\begin{remark}
  It can be observed that the linear velocity $v_x$ decreases to the lower bound $v_{\text{min}}$ only if the robot reaches the target configuration $\bm{g}_d$, where $\bm{p}=\bm{p}_d$ and $\theta=\theta_d=\theta_r$.
  For robots such as wheeled UGV, $v_{\text{min}}$ could be chosen as zero so that the robot could asymptotically converge to $\bm{g}_d$. As for fixed wing UAVs and other robots with $v_{\text{min}}>0$, they are not able to converge to a specific configuration and thereby the result of convergence to the target positive limit set and repeated arrival at the target configuration is the best we can achieve.
\end{remark}

When the robot's orientation $\theta$ aligns with the reference orientation $\theta_r$, the robot's trajectory becomes the time reparameterization of the integral curves of the CVF under the proposed linear velocity, referring to \cite[Chapter 1.5]{chicone2006ordinary}.
This implies that the robot is able to track the CVF and converge to the target positive limit set by aligning its orientation with $\theta_r$ under the angular velocity:
\begin{equation}\label{proportional}
  \omega_0 = -k_\omega\theta_e+\omega_r.
\end{equation}
In the above equation, $-k_\omega\theta_e$ is the feedback term and the feedforward term, i.e., the time derivative of $\theta_r$, is given by
\begin{equation}\label{ref_kine}
    \omega_r=\dot{\theta}_r=A(r_\delta)v_x\cos{\Delta \theta},
\end{equation}
where $\Delta \theta = \theta - \theta_\nabla$, $\theta_\nabla$ and $A(r_\delta)$ denote the direction and magnitude of the gradient $\nabla\theta_r$ of reference orientation, respectively.
The detailed derivations of \eqref{ref_kine} are provided in Appendix~\ref{appendix2}.

To ensure that the robot has curvature-bounded trajectory before it tracks the CVF, we refer to \eqref{curvature_constraint} and impose the angular velocity with a saturation bound given by
\[
  \bar{\omega}=v_x\bar{\kappa},
\]
and therefore, the saturated angular velocity is expressed as
\begin{equation}\label{omega}
  \omega = \left\{\begin{matrix}
\omega_0,
&\lvert\omega_0\rvert \le \bar{\omega}, \\
\bar{\omega}\text{sgn}(\omega_0),
& \lvert\omega_0\rvert > \bar{\omega},
\end{matrix}
  \right.
\end{equation}
where $\omega_0$ is determined by\eqref{proportional}, $\mathrm{sgn}(\cdot)$ is the sign function. 

According to \eqref{omega_r} and \eqref{g(r)} in Appendix~\ref{appendix2}, there holds that
\[
  \sup{\|\omega_r(\bm{g})\|} > \bar{\omega},\ \forall r_\delta < \rho.
\]
This equation implies that the upper bound of the feedforward term is larger than the saturation bound when $\bm{g}\in\mathcal{S}=\{\bm{g} \mid \|\bm{p}-\bm{p}_\delta\|<\rho\}$. Therefore, the angular velocity \eqref{omega} possibly saturates in $\mathcal{S}$ even when the feedback term is omitted. In other words, $\mathcal{S}$ is the minimal region where saturation occurs with the presence of feedback term.

To stabilize the orientation error $\theta_e$, we aim to establish the condition under which the angular velocity \eqref{omega} does not persistently saturate. In this way, $\theta_e$ is ultimately stabilized by $\omega=\omega_0$. From a geometric perspective, the robot's trajectory, whose curvature is bounded by $\bar{\kappa}$, cannot be enclosed in an open circular area with radius $\rho=1/\bar{\kappa}$ such as $\mathcal{S}$. Enlightened by this, we ensure that \eqref{omega} does not saturate persistently by confining the region where saturation occurs to the minimal extent of $\mathcal{S}$, as elaborated in the following theorem.

\begin{theorem}[Dynamic gain]\label{thm dynamic gain}
  The angular velocity only saturates in the region $\mathcal{S}=\{\bm{g} \mid \|\bm{p}-\bm{p}_\delta\|<\rho\}$, when the gain in \eqref{proportional} is specified by
  \begin{equation}\label{dynamic gain}
      k_\omega(\bm{g}) = \min{\left\{ \bar{k}_\omega, \frac{v_x}{\lvert \theta_e \rvert}(\bar{\kappa}-k(r_\delta)\lvert\cos{\Delta \theta}\rvert)\right\}},
  \end{equation}
  where $\bar{k}_\omega$ is the user-defined maximum gain and $k(r_\delta)$ is a continuous nonlinear function satisfying
  \begin{subequations}\label{gain condition}
      \begin{align}
          &0 < k(r_\delta) \le \bar{\kappa},\ \forall r_\delta >0, \label{k<kappa}\\
      &k(r_\delta) \ge A(r_\delta), \ \forall r_\delta \ge \rho. \label{k}
      \end{align}
  \end{subequations}
\end{theorem}
\begin{IEEEproof}
  See Appendix~\ref{appendix_S}.
\end{IEEEproof}

\begin{remark} \label{remark:dynamic gain}
    The proposition that the angular velocity does not saturate persistently can be verified with a counter example. 
    Assume that the angular velocity saturates all the time, which requires the robot to stay in $\mathcal{S}$. Since the target configuration $\bm{g}_d\notin \mathcal{S}$, the linear velocity satisfies $v_x>v_{min}\ge0$. In this case, the robot moves along a circular arc with radius $\rho$ since $\left|\omega\right|=\bar{\omega}$ and $\kappa=\left|\bar{\omega}/v_x\right|=\bar{\kappa}=1/\rho$. However, such a circle cannot be contained within $\mathcal{S}$, as $\mathcal{S}$ is itself an open circular region with radius $\rho$. This implies that the robot will leave $\mathcal{S}$ even when its angular velocity persistently saturates, which contradicts with Theorem~\ref{thm dynamic gain}. Above discussions further imply that the robot leaves the region $\mathcal{S}$ and the saturation only occurs in finite time.
\end{remark}

Due to the geometry of the CVF, i.e., the magnitude $A(r_\delta)$ of gradient $\nabla\theta_r$ exceeds the maximum curvature around the singular point $\bm{p}_\delta$. Therefore, the saturation of the angular velocity within $\mathcal{S}$ is inevitable. 
To ensure that the robot leaves such region and hence the angular velocity only saturates in finite time, we propose the dynamic gain in the above theorem. However, this dynamic gain vanished when the robot circulates on the boundary of $\mathcal{S}$, i.e., $k_\omega=0$, $\forall \bm{g} \in \mathcal{N}_{k_\omega}=\left\{\bm{g} \mid r_\delta = \rho, \lvert \theta_e \rvert = \pi/2 \right\}$.
In addition, due to the singularity of the CVF and the definition of orientation error, there are two singular sets for the robot, i.e., $\mathcal{N}_{\mathbf{T}}=\left\{\bm{g} \mid \bm{p}=\bm{p}_\delta \right\}$ and $\mathcal{N}_{\theta}=\left\{\bm{g} \mid \lvert \theta_e \rvert =\pi \right\}$. The above singular sets together form the non-converging set $\mathcal{N}=\mathcal{N}_{\mathbf{T}} \cup \mathcal{N}_{\theta} \cup \mathcal{N}_{k_\omega}$ of the closed-loop system.

It is worth noting that the dynamic gain not only responds to the change in CVF under robot's motion, it also imposes the inherent constraint $A(r_\delta)\le\bar{\kappa}$, $\forall r_\delta \ge \rho$ on the geometric properties of CVF.
To address the above intrinsic coupling between the CVF and control laws under the curvature constraint and to ensure the existence of dynamic gain, the co-design of the CVF and control laws is of great necessity. 
Based on the flexible adjustment of the CVF's parameters under \eqref{curvature condition}, we establish the condition that ensures the orientation error is stabilized by the saturated control laws with dynamic gain in the following theorem.
\begin{theorem}[Stabilization Condition of Orientation Error] \label{thm stabilization condition}
  For the closed-loop system \eqref{nonholonomic kinematics}, \eqref{curvature_constraint}, \eqref{vx}, \eqref{omega}, the orientation error $\theta_e$ is monotonically stabilized except for initial condition in the zero measure non-converging set $\mathcal{N}=\mathcal{N}_{\mathbf{T}} \cup \mathcal{N}_{\theta} \cup \mathcal{N}_{k_\omega}$ when the parameters of CVF satisfy
  \begin{equation}\label{stabilization condition}
    \frac{1}{r_{i-1}} + \frac{1}{r_{i}-r_{i-1}} \le \bar{\kappa}, \quad i=2,3.
  \end{equation}
\end{theorem}
\begin{IEEEproof}
  See Appendix~\ref{appendix_theta_e}.
\end{IEEEproof}

\begin{remark}
  In \cite{yao2023guiding}, the angular velocity for Dubins-car-like robots to track the VF is also subject to saturation. 
  However, the orientation error is shown to be stabilized only when the angular velocity does not saturate persistently. To achieve this, implicit constraints on the system dynamics are introduced. However, it remains unclear how to establish these constraints in practice. 
  In contrast, our approach offers a straightforward criterion for the co-design of the CVF and control laws. Under the proposed control inputs, the nonholonomic robot exits the saturation region under curvature constraint, and hence the angular velocity would not saturate persistently, leading to the results that the orientation error is monotonically stabilized.
\end{remark}

According to the decomposition of workspace in \eqref{decomposition}, the first term on the left-hand side of \eqref{stabilization condition} reflects the distance between the singular point $\bm{p}_\delta$ and the region $\mathcal{A}_{i=2,3}$, where the flow fields are blended. The second term depicts the range of the blending region $\mathcal{A}_{i=2,3}$. To ensure the existence of dynamic gain that minimizes the saturation region to $\mathcal{S}$, the changing rate in CVF's orientation should be limited by the saturation bound $\bar{\omega}$ outside $\mathcal{S}$. When the blending region is close to the singular point and hence the CVF changes rapidly and the first term is large, the range of the blending region should be enlarged to reduce the effect of singularity, which requires a smaller second term.

Through the co-design of the CVF and control laws, the orientation error is stabilized as stated in Theorem~\ref{thm stabilization condition}. In this way, the nonholonomic robot tracks the CVF without violating the curvature constraint. Building on these results, we further utilize input-to-state stability (ISS) to demonstrate the almost global convergence of the robot's configuration in the following theorem.

\begin{theorem}[Convergence to Target Positive Limit Set] \label{thm convergence}
    With the control inputs \eqref{vx} and \eqref{omega}, the nonholonomic robot \eqref{nonholonomic kinematics} subject to curvature constraint \eqref{curvature_constraint} converges to the target positive limit set $\mathcal{L}_d^+(\bm{g}_0) \ni \bm{g}_d$, except for initial configuration in the zero measure non-converging set $\mathcal{N}$.
\end{theorem}

\begin{IEEEproof}
  See Appendix~\ref{appendix_convergence}.
\end{IEEEproof}

\begin{remark} \label{remark:comparison}
  In the literature, most VF-based motion planning algorithms for nonholonomic robots with curvature constraints focus on incorporating the curvature constraint either in the design of the VF \cite{lau2015fluid,pothen2017curvature,shivam2021curvature} or in the derivation of control laws \cite{yao2023guiding}. Consequently, it remains unclear whether these robots can reliably follow the VFs under limited actuation. We fill this gap in the paper by integrating the curvature constraint into the co-design of the CVF and saturated control laws with dynamic gain, providing rigorous guarantee on the convergence of the nonholonomic robot under curvature constraint.
\end{remark}

\begin{figure*}[htbp]
  \centering
  \subfigure[Exp 1, time in $\mathcal{S}$ emphasized on the right.]{\includegraphics[width=0.3\linewidth]{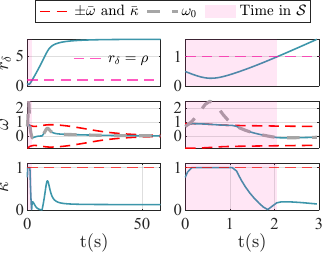}\label{fig:Lemma2_Exp1}}
  \hfill
  \subfigure[Exp 2, time in $\mathcal{S}$ emphasized on the right.]{\includegraphics[width=0.3\linewidth]{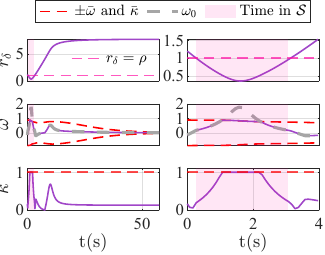} \label{fig:Lemma2_Exp2}}
  \hfill
  \subfigure[Exp 3, time in $\mathcal{S}$ emphasized on the right.]{\includegraphics[width=0.3\linewidth]{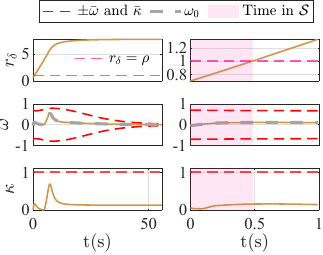} \label{fig:Lemma2_Exp3}}
  \subfigure[Exp 4 without saturation.]{\includegraphics[width=0.23\linewidth]{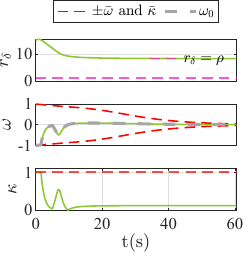} \label{fig:Lemma2_Exp4}}
  \hfill
  \subfigure[Exp 5 without saturation.]{\includegraphics[width=0.23\linewidth]{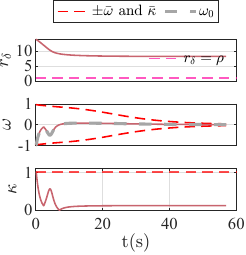} \label{fig:Lemma2_Exp5}}
  \hfill
  \subfigure[Exp 6 without saturation.]{\includegraphics[width=0.23\linewidth]{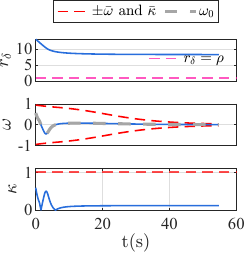} \label{fig:Lemma2_Exp6}}
  \hfill
  \subfigure[Exp 7 without saturation.]{\includegraphics[width=0.23\linewidth]{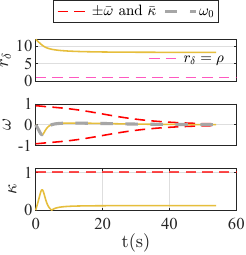} \label{fig:Lemma2_Exp7}}
  \caption{Simulation results that verify Theorem~\ref{thm dynamic gain}. The subfigures~\ref{sub@fig:Lemma2_Exp1}-\ref{sub@fig:Lemma2_Exp3} present the results for Exp 1-3 in the first set. The time intervals during which the robot is moving in the saturation region $\mathcal{S}$ are highlighted with magenta shaded area. The entire simulation results are on the left and the shaded areas are magnified on the right. The results of the Exp 4-7 in the second set are shown by subfigures~\ref{fig:Lemma2_Exp4}-\ref{sub@fig:Lemma2_Exp7}.}
  \label{fig:Lemma 2}
\end{figure*}

\begin{figure}[!hbtp]
  \centering
  \includegraphics[width=0.9\linewidth]{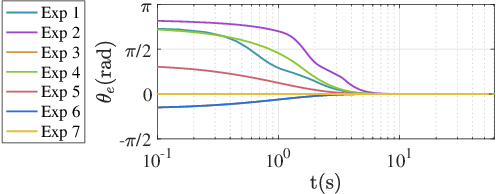}
  \caption{The orientation error $\theta_e$ of each example monotonically converges to zero, confirming Theorem~\ref{thm stabilization condition}. The time is in logarithmic scale.}
  \label{fig:Thm3}
\end{figure}

\begin{figure*}[!hbtp]
  \centering
  \subfigure[Configuration errors and control inputs for Exp 1-3.]{\includegraphics[width=0.45\linewidth]{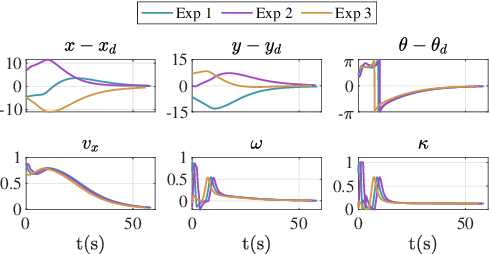} \label{fig:Thm4 Set 1 error}} \hfill
  \subfigure[Configuration errors and control inputs for Exp 4-7.]{\includegraphics[width=0.45\linewidth]{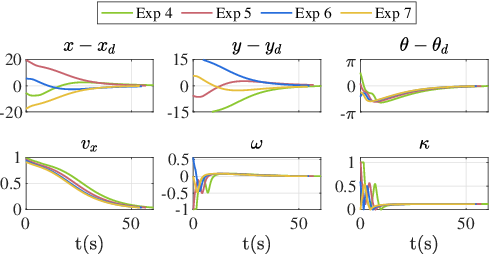} \label{fig:Thm4 Set 2 error}}

  \subfigure[Snapshot at $t=0$.]{\includegraphics[width=0.23\linewidth]{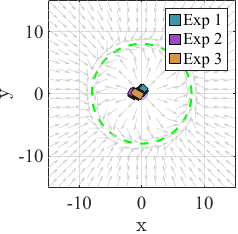} \label{fig:Thm4 Set 1 t0}} \hfill
  \subfigure[Snapshot at $t_i=t_{i,f}/3$.]{\includegraphics[width=0.23\linewidth]{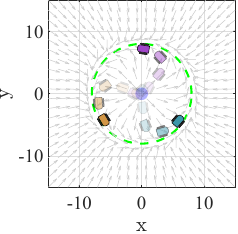} \label{fig:Thm4 Set 1 t1}} \hfill
  \subfigure[Snapshot at $t_i=2t_{i,f}/3$.]{\includegraphics[width=0.23\linewidth]{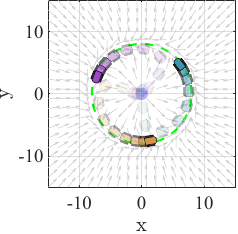} \label{fig:Thm4 Set 1 t2}} \hfill
  \subfigure[Snapshot at $t_i=t_{i,f}$.]{\includegraphics[width=0.23\linewidth]{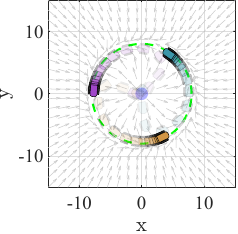} \label{fig:Thm4 Set 1 t3}}

  \subfigure[Snapshot at $t=0$.]{\includegraphics[width=0.23\linewidth]{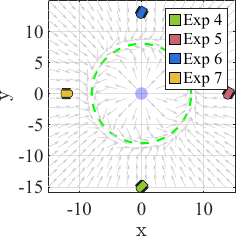} \label{fig:Thm4 Set 2 t0}} \hfill
  \subfigure[Snapshot at $t_i=t_{i,f}/3$.]{\includegraphics[width=0.23\linewidth]{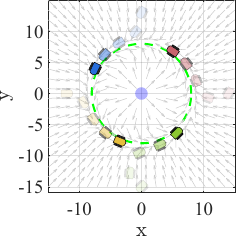} \label{fig:Thm4 Set 2 t1}} \hfill
  \subfigure[Snapshot at $t_i=2t_{i,f}/3$.]{\includegraphics[width=0.23\linewidth]{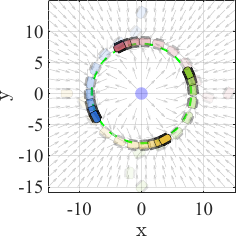} \label{fig:Thm4 Set 2 t2}} \hfill
  \subfigure[Snapshot at $t_i=t_{i,f}$.]{\includegraphics[width=0.23\linewidth]{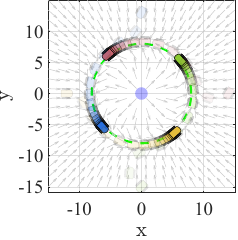} \label{fig:Thm4 Set 2 t3}}
  \caption{Simulation results validate Theorem~\ref{thm convergence}. Denote the simulation time for the $i$-th example as $t_i$ and the terminal time as $t_{i,f}$. The subfigures~\ref{sub@fig:Thm4 Set 1 error},~\ref{sub@fig:Thm4 Set 2 error} show that the configuration in each example converges to the target configuration $\bm{g}_d=(x_d,y_d,\theta_d)$ under the control inputs. The subfigures~\ref{sub@fig:Thm4 Set 1 t0}-\ref{sub@fig:Thm4 Set 1 t3} present snapshots of Example 1-3 at different time instances. The subfigures~\ref{sub@fig:Thm4 Set 2 t0}-\ref{sub@fig:Thm4 Set 2 t3} are snapshots of Example 4-7 at different time instances.}
  \label{fig:Thm 4}
\end{figure*}

\section{Numerical Simulations}\label{sec:numerical simulation}
In this section, we verify the results in Section~\ref{sec:control} with numerical simulation examples and compare the proposed method with several existing VF-based motion planning algorithms via Monte Carlo experiments.

\subsection{Theoretical Result Verification} \label{subsec:verification}

\subsubsection{Setup}
To validate our theoretical results in Section~\ref{sec:control}, we conduct numerical simulations and investigate the behavior of the nonholonomic robot under the proposed framework.

In the simulation, the nonholonomic robot has the minimum turning radius $\rho = 1$.
The design parameters for CVF satisfying \eqref{curvature condition} and \eqref{stabilization condition} are specified by $r_1=4\rho$, $r_2=8\rho$ and $r_3=12\rho$.
For the linear velocity control law \eqref{vx}, it is bounded by $v_{\text{min}}=0$, $v_{\text{max}}=1$, and the parameters are set as $c_{\bm{p}}=12\rho$, $c_\theta=\pi$.
The nonlinear function $k(\cdot)$ in the dynamic gain \eqref{dynamic gain} is chosen as
\begin{equation} \label{chosen k}
  k(r_\delta)=\left\{
\begin{matrix}
  & r_\delta/\rho^2 & r_\delta < \rho,\\
  & 1/r_\delta + \theta_{r,r_\delta} & r_\delta \ge \rho,
\end{matrix}\right.
\end{equation}
where $\theta_{r,r_\delta}$ is given in \eqref{g(r)} and the maximum gain is $\bar{k}_\omega=1$.

We present two sets of simulation examples, each of which is conducted independently.
In the first set, we prepare three examples in which the robot starts in or passes through the saturation region $\mathcal{S}=\left\{\bm{g} \mid r_\delta = \| \bm{p}-\bm{p}_\delta \| < \rho\right\}$. The second set includes four examples where the robot stays outside the region $\mathcal{S}$.
The initial and target configurations for each example are summarized in Table~\ref{tab:validation}.

\begin{table}[!hbtp]
  \centering
  \caption{Initial and Target Configurations for Simulations}
  \begin{tabular}{| c | c | c c c | c c c |}
  \hline
    \multicolumn{2}{|c|}{\textbf{Examples}} & \multicolumn{3}{c|}{$\bm{g}_0=(x_0,y_0,\theta_0)$} & \multicolumn{3}{c|}{$\bm{g}_d=(x_d,y_d,\theta_d)$} \\
  \hline
  \multirow{3}{*}{\textbf{Set 1}}
   & \textbf{Exp.1} & (0 & 0.5 & 5$\pi$/4)  & (4 & 4$\sqrt{3}$ & 5$\pi$/6) \\  \cline{2-8}
   & \textbf{Exp.2} & (-1.2 & 0 & -$\pi$/6) & (-8 & 0 & -$\pi$/2) \\ \cline{2-8}
   & \textbf{Exp.3} & (-0.7 & 0 & 5$\pi$/6) & (4 & -4$\sqrt{3}$ & $\pi$/6) \\ \hline
   \multirow{4}{*}{\textbf{Set 2}}
   & \textbf{Exp.4} & (0 & -15 & 5$\pi$/4)  & (4$\sqrt{2}$ & 4$\sqrt{2}$ & 3$\pi$/4) \\ \cline{2-8}
   & \textbf{Exp.5} & (14 & 0 & -2$\pi$/3)  & (-4$\sqrt{2}$ & 4$\sqrt{2}$ & -3$\pi$/4) \\ \cline{2-8}
   & \textbf{Exp.6} & (0 & 13 & -2$\pi$/3)  & (-4$\sqrt{2}$ & -4$\sqrt{2}$ & -$\pi$/4) \\ \cline{2-8}
   & \textbf{Exp.7} & (-12 & 0 & 0)         & (4$\sqrt{2}$ & -4$\sqrt{2}$ & $\pi$/4) \\ \cline{2-8}
  \hline
  \end{tabular}
  \label{tab:validation}
\end{table}

\subsubsection{Results}

We begin with the saturation behavior depicted in Theorem~\ref{thm dynamic gain}.
In Figures~\ref{fig:Lemma2_Exp1} and~\ref{fig:Lemma2_Exp2}, the unsaturated angular velocity $\omega_0$ in \eqref{proportional} is represented by the gray dotted line. It exceeds the bound $\bar{\omega}$ presented by the red dotted line, thereby the angular velocity saturates when the robot is in the saturation region $\mathcal{S}=\left\{ \bm{g} \mid r_\delta < \rho \right\}$.
In contrast, in Figures~\ref{fig:Lemma2_Exp3}-\ref{sub@fig:Lemma2_Exp7}, the condition $\omega=\omega_0$ always holds, implying that the angular velocity does not saturate. 
Moreover, the dynamic gain also enables the robot to leave the saturation region in finite time such that the angular velocity would not saturation persistently, as shown by Figures~\ref{fig:Lemma2_Exp1}-\ref{fig:Lemma2_Exp3}.

Following the above discussion, we illustrate the evolution of the orientation error $\theta_e$ in Figure~\ref{fig:Thm3}.
By co-designing of the CVF and control laws, the orientation error $\theta_e$ asymptotically converges to zero in all examples as stated in Theorem~\ref{thm stabilization condition}.
It should be noted that the initial orientation of the robot in Example 7 aligns with the CVF, such that the orientation error $\theta_e$ remains zero.

Finally, Figure~\ref{fig:Thm 4} shows the results of almost global convergence in Theorem~\ref{thm convergence}.
Under the proposed motion planning algorithm, the configuration error of the nonholonomic robot converges to zero while the curvature of the actual trajectory maintains the prescribed upper bound $\bar{\kappa}=1$, as illustrated in Figures~\ref{fig:Thm4 Set 1 error},\ref{sub@fig:Thm4 Set 2 error}.
The snapshots of trajectories shown in Figures~\ref{fig:Thm4 Set 1 t0}-\ref{sub@fig:Thm4 Set 1 t3} and Figures~\ref{fig:Thm4 Set 2 t0}-\ref{sub@fig:Thm4 Set 2 t3} depict the convergence towards the target configuration for all examples.

\subsection{Comparative Study} \label{subsec:comparative}

\subsubsection{Setup}
In this subsection, we conduct 1000 trials of Monte Carlo experiments with different initial and target configurations to compare the performance of the proposed CVF method, with respect to other VF-based motion planning algorithms with feedback control laws for nonholonomic robots. The benchmarks include the dynamic vector field (DVF) \cite[Eq.20, Eq.24]{he2024simultaneous}, the dipole-like attractive vector field (AVF) \cite[Eq.8, Eq.18]{Panagou2017Distributed}, the curvature-constrained Lyapunov vector field (CLVF) \cite[Eq.16 Eq.50]{pothen2017curvature}, and the guiding vector field (GVF) \cite[Eq.4, Eq.24]{yao2021singularity}.
Among these candidates, the DVF and AVF guide the unicycle to the target configuration, while the CLVF and GVF enable the UAV to follow a target circular path, with both scenarios falling within the scope of the proposed approach. The comparisons are categorized into two groups: comparisons between the DVF and AVF on the unicycle, and comparisons between the GVF and CLVF on the UAV.

For the unicycle with linear velocity bounded by $v_{\text{max}}=3$ and $v_{\text{min}}=0$, we prepare four sets of target configurations, where the target positions are uniformly distributed on the circle $x^2+y^2=r_2^2$ and the target orientation is tangent to the circle.
For UAV, the linear velocity is selected to be constant at $v_{\text{min}}=3$ and the target positive limit set is specified by $\mathcal{L}_d^+=\left\{ \|\bm{p}\|=r_2, \theta=\varphi+\pi/2 \right\}$, where four sets of target configurations are also uniformly distributed.
The maximum curvature is defined as $\bar{\kappa}=1$, the minimal turning radius as $\rho=1$, and the parameters of the proposed method remain consistent with those in the previous subsection.
We conduct 250 trials of Monte Carlo simulation for each target configuration, totaling 1000 trials for each candidate, with the initial configuration randomly distributed over $x_0\in [-15\rho, 15\rho]$, $y_0\in [-15\rho, 15\rho]$, and $\theta_0\in[0,2\pi)$.

\subsubsection{Metrics}
The metrics are divided into two categories to compare the reference trajectory quality and the control performance, respectively.
To evaluate the quality of the reference trajectories, we compute the percentage of trials in which the integral curves have curvature bounded by $\bar{\kappa}$. In addition, the relative length of the integral curves is defined as $L_r=L/\|\bm{p}_d-\bm{p}_0\|$, where $L$ is the length of the integral curve from initial position $\bm{p}_0$ to the target position $\bm{p}_d$.
Regarding the control performance, the percentage of trials where the control inputs satisfy the curvature constraint \eqref{curvature_constraint}.
Next, the average curvature of the trajectories over time is computed to reflect the feasibility of the control inputs w.r.t. the capability of the nonholonomic robot.
Moreover, we record the average time required for the robot to converge to the target configuration.
Finally, we define the root-mean-square error (RMSE) between $\{\omega(t_{i})\}$ and $\{\omega(t_{i+1})\}$ to measure the acceleration of control inputs, where $t_i$ is the $i$-th time step in the simulation.

\begin{table*}[!hptb]
  \centering
  \caption{Comparison Results across Vector Field-Based Algorithms, CVF Proposed in This Paper.}
  \begin{tabular}{|c|c|c|c|c|c|c|c|}
    \hline
    \multicolumn{2}{|c|}{\textbf{Metrics}} & \textbf{CVF (Unicycle)} & \textbf{DVF \cite{he2024simultaneous}} & \textbf{AVF \cite{Panagou2017Distributed}} & \textbf{CVF (UAV)} & \textbf{CLVF \cite{pothen2017curvature}} & \textbf{GVF \cite{yao2021singularity}} \\ \hline
    \multirow{2}{*}{\textbf{Trajectory Quality}} & Percentage of $\kappa\le \bar{\kappa}$ & \textbf{1.0000} & \textbackslash & 0.9960 & \textbf{1.0000} & \textbf{1.0000} & 0.8050 \\ \cline{2-8}
    & Relative length $L_r$ & 4.1448 & \textbackslash & 4.5976 & 4.1448  &  3.7906  &  \textbf{3.1228} \\ \hline
    \multirow{4}{*}{\textbf{Control Performance}} & Percentage of $\lvert \omega \rvert / \lvert v_x \rvert \le \bar{\kappa}$ & \textbf{1.0000} & 0.5330 & 0.9080 & \textbf{1.0000} & 0.0050 & 0.5840 \\ \cline{2-8}
     & Average curvature & \textbf{0.1415} & 0.2144 & 0.1902 & \textbf{0.1415} & 1.0350 & 0.1737 \\ \cline{2-8}
     & Average time & 28.5228 & 50.2299 & \textbf{19.9667} & 28.1643 & 14.6227 & \textbf{11.5851} \\ \cline{2-8}
     & $\omega_\text{RSME}$ & 0.0589 & \textbf{0.0081} & 0.0513 & \textbf{0.0587} & 6.7624 & 1468.9937 \\ \hline
    \end{tabular}
  \label{tab:comparison}
\end{table*}

\subsubsection{Results}

The comparison results are presented in Table~\ref{tab:comparison}. It can be observed that the proposed method outperforms the other algorithms by generating control inputs that adhere to the curvature constraint across all trials.

For the results of the unicycle, the DVF is not taken into account for the comparison of trajectory quality since it is a map from $\mathbb{R}^2\times \mathbb{S}$ to $\mathbb{R}^2$ and hence the integral curves do not represent the reference trajectories we consider in this paper. Throughout 1000 trials, the control inputs provided by DVF and AVF contain sharp turns that exceed the capability of the unicycle since the curvature constraint is not addressed in \cite{he2024simultaneous,Panagou2017Distributed}. In contrast, the proposed CVF-based algorithm provides reference trajectories with control inputs that both satisfy the curvature constraint under the guarantees established in Theorems~\ref{thm T} and~\ref{thm convergence}.

In the simulation based on fixed-wing UAV, only 80.5$\%$ of the reference trajectories provided by GVF exhibit bounded curvature. The percentage of actual trajectories with bounded curvature drops to 58.4$\%$ since the actual trajectories may deviate while tracking the GVF.
Moreover, the control inputs of the GVF become singular when the UAV approaches the singular point, which is reflected by the high value of $\omega_{\text{RSME}}$.
It is worth noting that the CLVF solely integrates the curvature constraint into the design of VF, and generates reference trajectories with curvature bounded by $\bar{\kappa}$ as in this paper. However, only 0.5$\%$ of the actual trajectories guided by CLVF can achieve the desired curvature bound, since it requires overly large angular velocity to track the CLVF when deviation occurs.
The above comparative study reveals the necessity and also the challenge of the co-design of the VF and control laws as considered in this paper. According to the results of the proposed algorithm, both the reference and actual trajectories are provably curvature-bounded in all trials.
As a trade-off between reliability and efficiency, the proposed method requires relatively longer time to converge to the target positive limit set.
Nevertheless, the proposed method successfully guarantees the curvature constraint and generates the lowest acceleration during convergence, as indicated by the lowest value of $\omega_{\text{RSME}}$.

\section{Hardware Experiments}\label{sec:Experiments}

As mentioned in Section~\ref{sec:preliminaries}, the proposed method under the problem formulation is applicable to a wide range of nonholonomic robots. In the real-world implementations presented in this section, we utilize two distinct robotic systems, namely the Ackermann UGV and the fixed-wing UAV, to further demonstrate the effectiveness of the proposed method.

\subsection{Experiments with the Ackermann UGV} \label{subsec:UGV}
\subsubsection{Setup}
\footnotetext[1]{https://github.com/agilexrobotics/limo-doc}
\footnotetext[2]{https://ymate3d.com}

\begin{figure}[hbtp]
  \centering
  \includegraphics[width=0.9\linewidth]{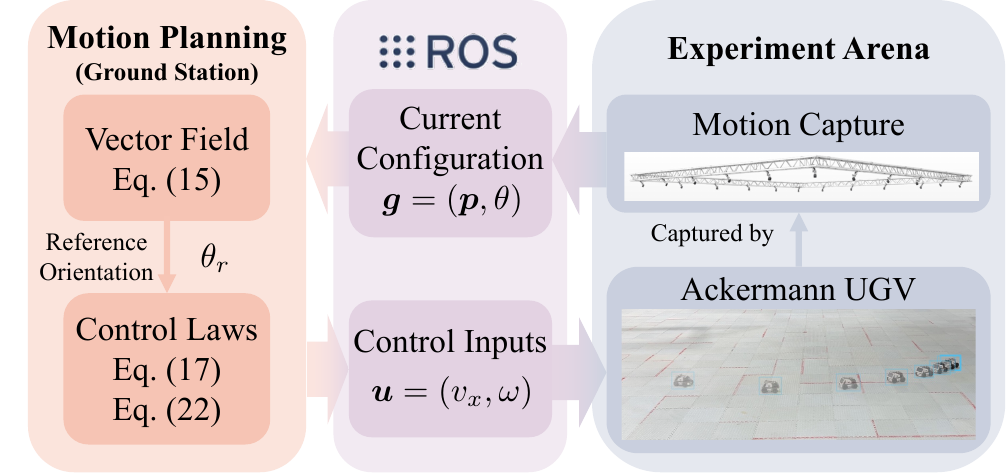}
  \caption{Experiment architecture for the Ackermann UGV.}
  \label{fig:Ackermann system}
\end{figure}

To verify the efficacy of the closed-loop motion planning algorithm in real-world applications, we select the Ackermann UGV \footnotemark[1] with minimum turning radius $\rho=\SI{0.6}{m}$ and maximum linear velocity $v_{\text{max}}=\SI{1}{m/s}$ to conduct experiments.
The experimental setup is illustrated in Figure~\ref{fig:Ackermann system}, where the ground station executes the motion planning algorithm, transmits real-time control inputs to the Ackermann UGV, and receives feedback on its configuration from FZMotion motion capture system \footnotemark[2] via the Robot Operating System (ROS).

Similar to the setup in the previous section, we prepare two sets of independently conducted examples for the Ackermann UGV.
In the first set of three examples, the UGV starts in or passes through the saturation region $\mathcal{S}$, while the UGV remains outside $\mathcal{S}$ for the four examples in the second set. The initial and target configurations are summarized in Table~\ref{tab:Ackermann}.
To generate control inputs that start from zero, we specify the linear velocity gain by $k_v = (1-e^{-0.3t})v_{\text{max}}$, where $t$ is the runtime and other control parameters are consistent with those in the previous section.

\begin{table}[!hbtp]
  \centering
  \caption{Initial and Target Configurations for Ackermann UGV}
  \begin{tabular}{| c | c | c c c | c c c |}
  \hline
    \multicolumn{2}{|c|}{\textbf{Examples}} & \multicolumn{3}{c|}{$\bm{g}_0=(x_0,y_0,\theta_0)$} & \multicolumn{3}{c|}{$\bm{g}_d=(x_d,y_d,\theta_d)$} \\
  \hline
  \multirow{3}{*}{\textbf{Set 1}}
   & \textbf{Exp.1} & (27.33 & -1.35 & 3.08)  & (29.12 & 2.81 & 2.62) \\  \cline{2-8}
   & \textbf{Exp.2} & (26.41 & -1.30 & -1.15) & (21.92 & -1.35 & -1.57) \\ \cline{2-8}
   & \textbf{Exp.3} & (26.59 & -1.14 & 1.68) & (29.12 & -5.51 & 0.52) \\ \hline
   \multirow{4}{*}{\textbf{Set 2}}
   & \textbf{Exp.4} & (32.86 & -7.50 & -1.24)  & (30.11 & 2.04 & 2.36) \\ \cline{2-8}
   & \textbf{Exp.5} & (32.57 & 4.59 & -0.03)  & (23.32 & 2.04 & -2.35) \\ \cline{2-8}
   & \textbf{Exp.6} & (21.43 & 2.36 & 0.48)  & (23.32 & -4.74 & -0.78) \\ \cline{2-8}
   & \textbf{Exp.7} & (19.96 & -4.99 & 0.80) & (30.11 & -4.74 & 0.78) \\ \cline{2-8}
  \hline
  \end{tabular}
  \label{tab:Ackermann}
\end{table}

\subsubsection{Results}
\begin{figure*}[!hbtp]
  \centering
  \subfigure[Control inputs tracking  of Exp 1-3.]{\includegraphics[height=0.45\linewidth]{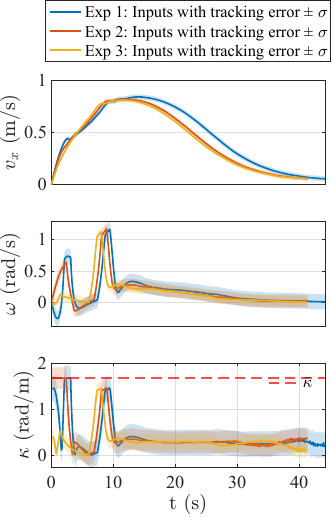} \label{fig:Exp_inputs_123}}
  \hfill
  \subfigure[Control inputs tracking results of Exp 4-7.]{\includegraphics[height=0.45\linewidth]{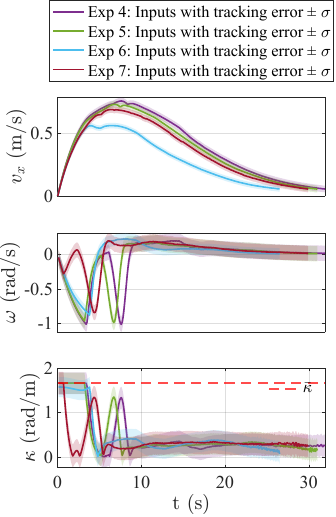} \label{fig:Exp_inputs4567}}
  \hfill
  \subfigure[Configuration errors.]{\includegraphics[height=0.45\linewidth]{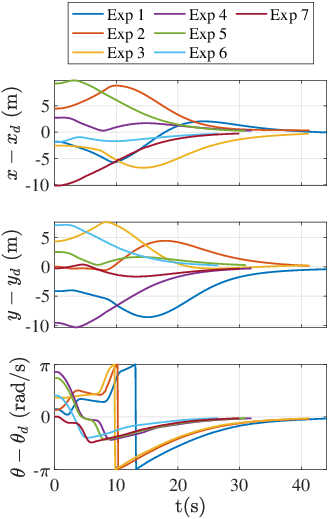}\label{fig:Exp_convergence}}

  \subfigure[Initial configurations for Exp 1-3.]{\includegraphics[width=0.48\linewidth]{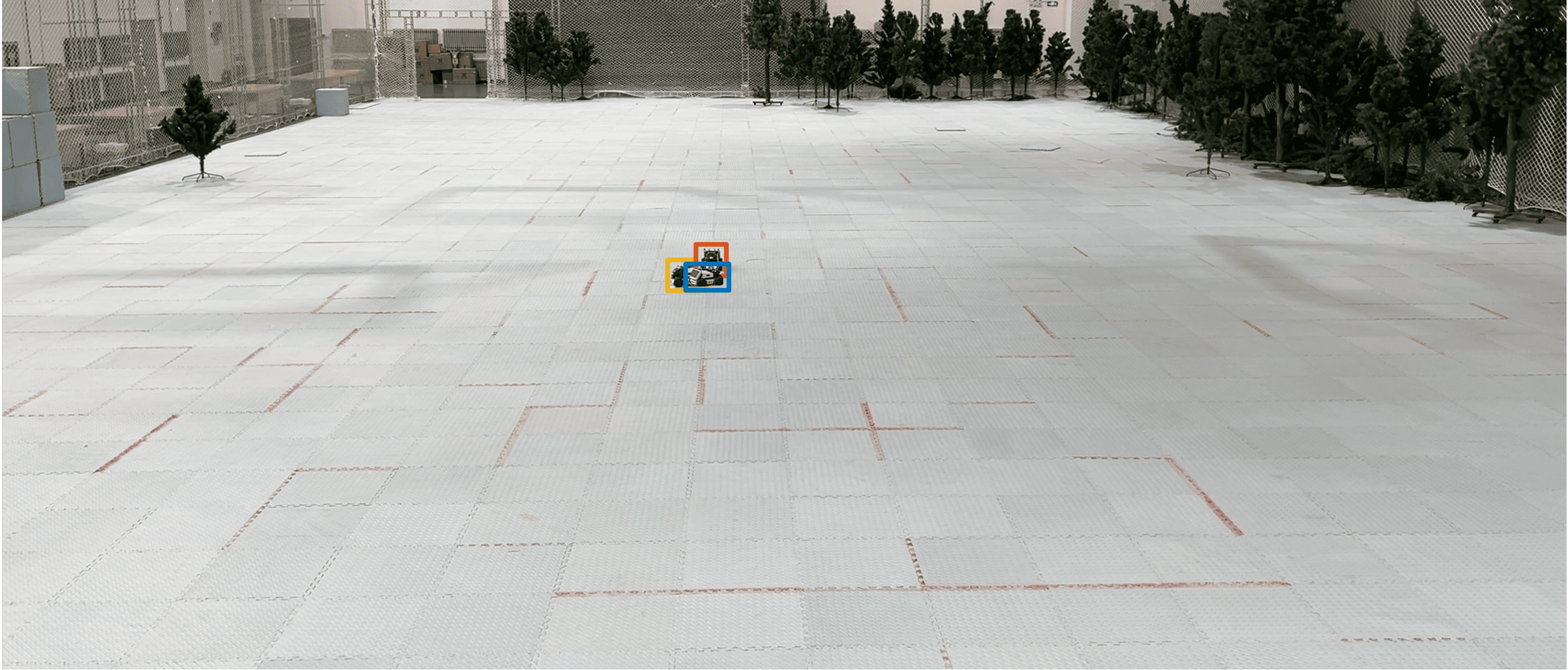} \label{fig:Exp_init_123}}
  \hfill
  \subfigure[Initial configurations for Exp 4-7.]{\includegraphics[width=0.48\linewidth]{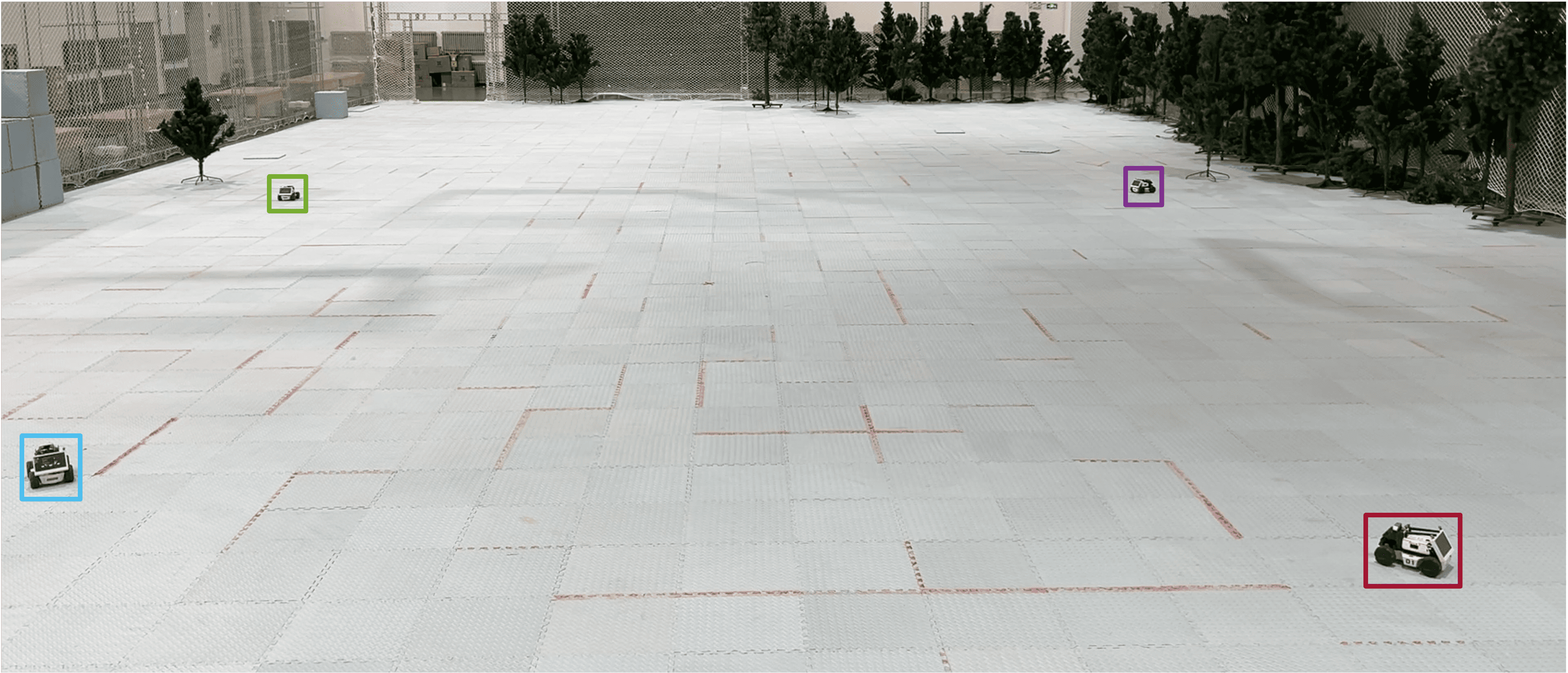} \label{fig:Exp_init_4567}}
  \hfill
  \subfigure[Halftime of Exp 1-3.]{\includegraphics[width=0.48\linewidth]{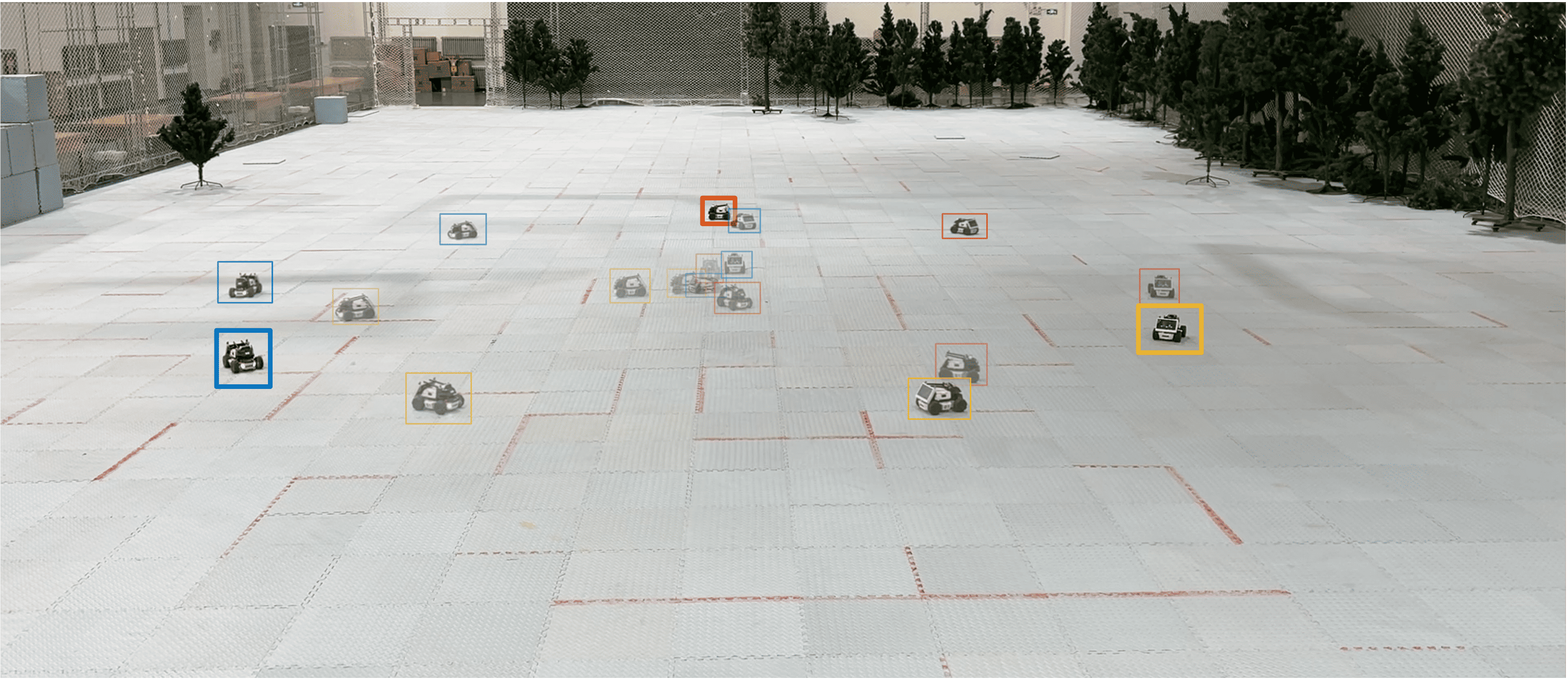} \label{fig:Exp_half_123}}
  \hfill
  \subfigure[Halftime of Exp 4-7.]{\includegraphics[width=0.48\linewidth]{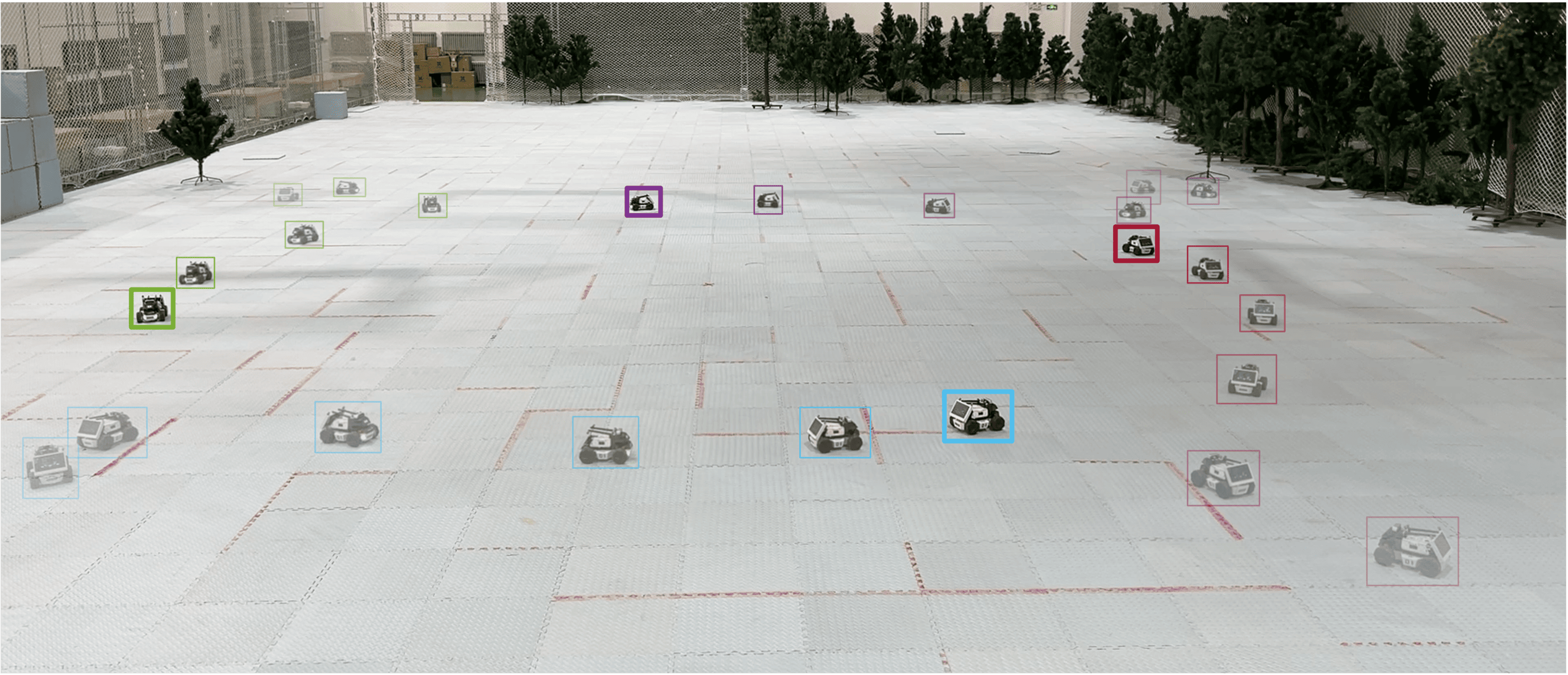} \label{fig:Exp_half_4567}}
  \hfill
  \subfigure[Full trajectories of Exp 1-3.]{\includegraphics[width=0.48\linewidth]{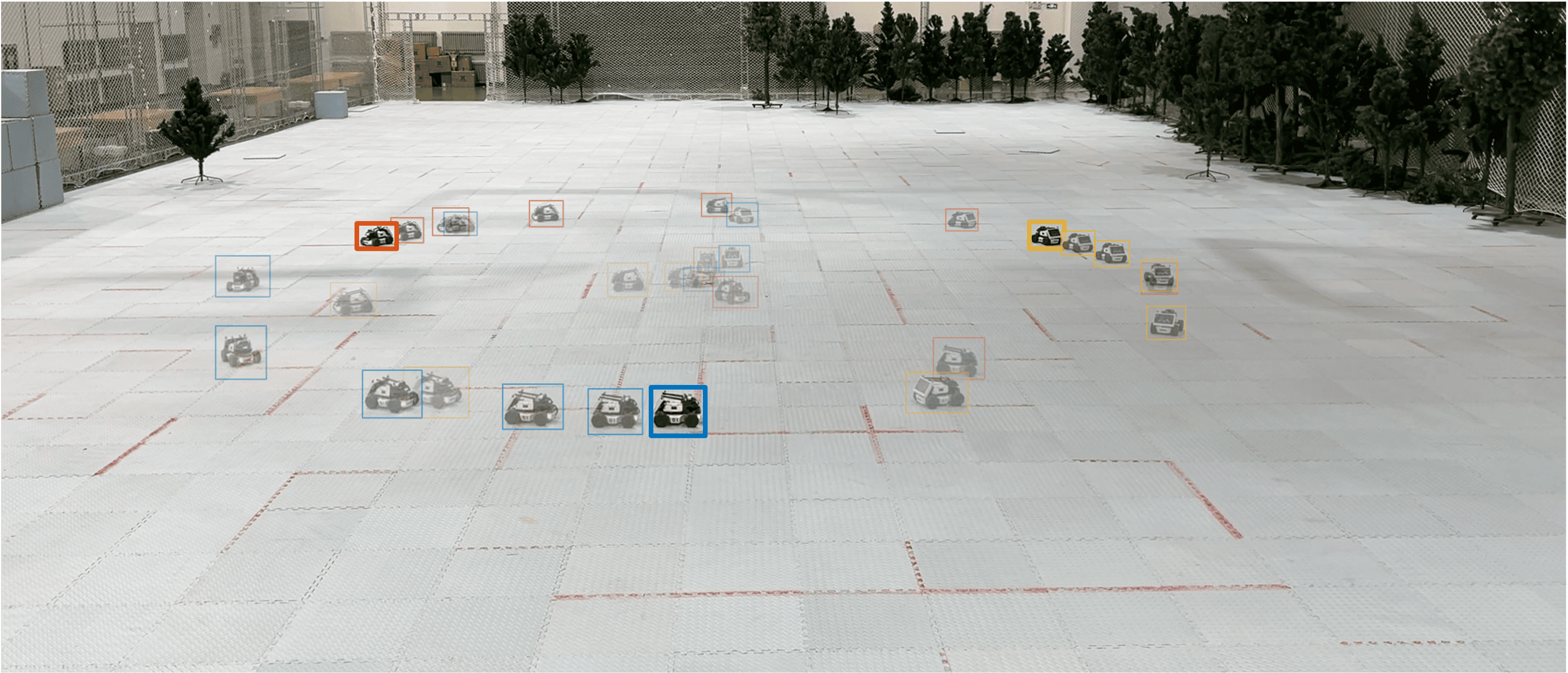} \label{fig:Exp_full_123}}
  \hfill
  \subfigure[Full trajectories of Exp 4-7.]{\includegraphics[width=0.48\linewidth]{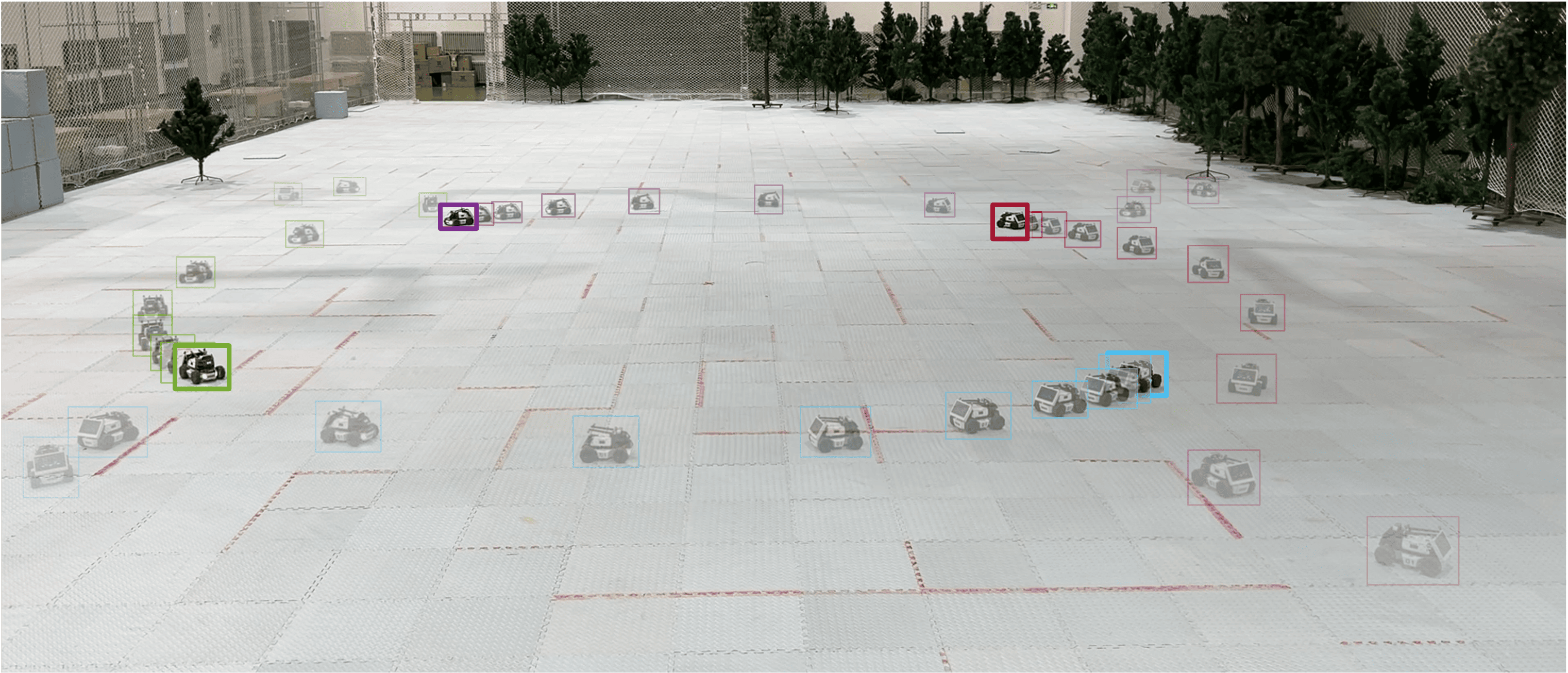} \label{fig:Exp_full_4567}}
  \caption{Experimental results with the Ackermann UGV. The tracking errors of the control inputs are presented by the shaded areas around the desired values in subfigures~\ref{sub@fig:Exp_inputs_123},~\ref{sub@fig:Exp_inputs4567}.}
  \label{fig:Ackermann}
\end{figure*}

The experimental results are presented in Figure~\ref{fig:Ackermann}.
Figures~\ref{fig:Exp_inputs_123} and~\ref{sub@fig:Exp_inputs4567} show the tracking results of the control inputs.
The errors between the current and target configurations are plotted in Figure~\ref{fig:Exp_convergence}.
The snapshots of Examples 1-3 and Examples 4-7 are presented in Figures~\ref{fig:Exp_init_123},~\ref{sub@fig:Exp_half_123},~\ref{sub@fig:Exp_full_123} and~\ref{fig:Exp_init_4567},~\ref{sub@fig:Exp_half_4567},~\ref{sub@fig:Exp_full_4567}, respectively.
Due to measurement noise, we observe that the tracking results of the linear velocity exhibit small fluctuations around the desired value. In the case of angular velocity, the tracking error increases due to parameter uncertainties and the response time of the UGV's chassis. However, despite the aforementioned real-world disturbances, the Ackermann UGV consistently converges to the target configuration with bounded trajectory curvature in all examples. The experimental results showcase the effectiveness and robustness of the feedback motion planning algorithm in real-world applications.

\subsection{HIL Experiments with the Fixed-Wing UAV} \label{subsec:HIL}
\begin{figure}[!hbtp]
  \centering
  \includegraphics[width=0.9\linewidth]{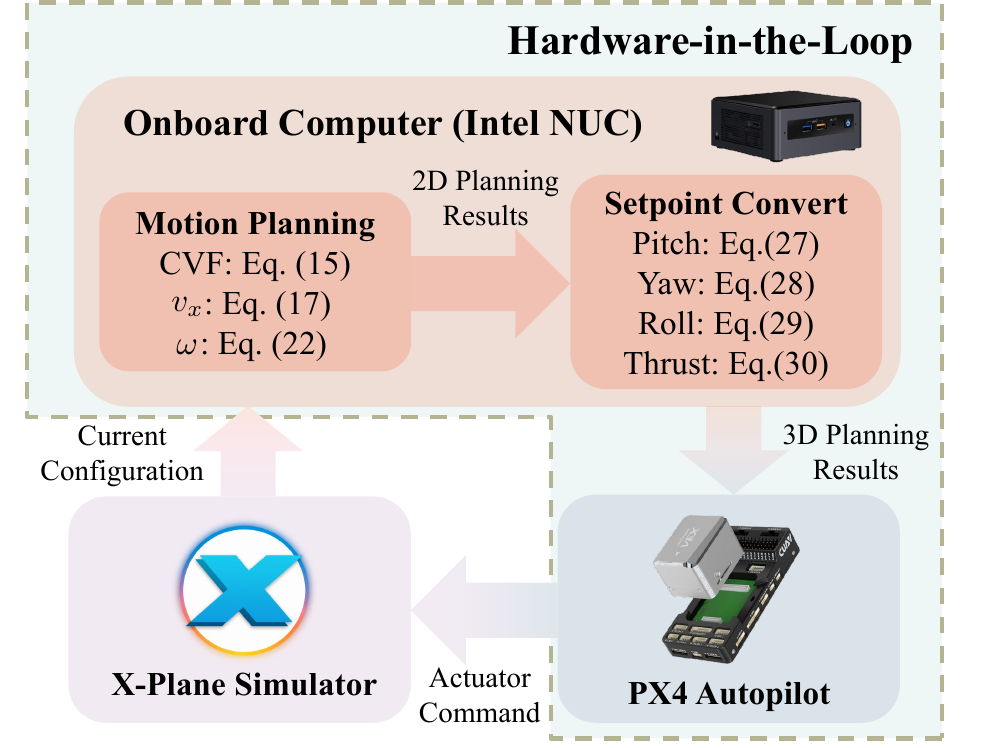}
  \caption{HIL experiment platform same as \cite{liu2024low} except the planning algorithm.}
  \label{fig:Semi-Physical}
\end{figure}

\begin{figure}[!hbtp]
  \centering
  \includegraphics[width=0.8\linewidth]{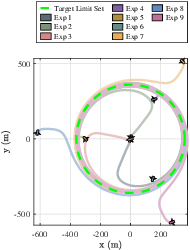}
  \caption{Initial configuration and trajectory for each example, where all the target configurations are on the same target positive limit set, illustrated by the green dotted circle.}
  \label{fig:UAV trajectories}
\end{figure}

\begin{figure*}[!hbtp]
  \centering
  \includegraphics[width=0.8\linewidth]{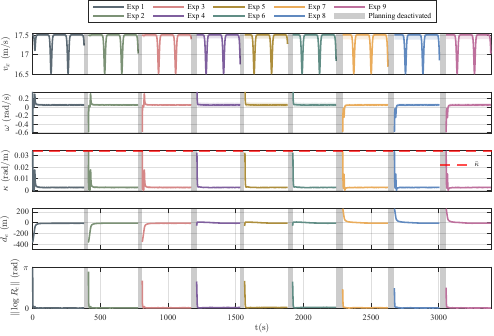}
  \caption{Results of HIL experiments with fixed-wing UAV, where the tracking errors $\sigma$ of $v_x$, $\omega$ and $\kappa$ are shown by shaded areas around the desired value. The intervals between two examples, where the motion planning algorithm is deactivated and the autopilot takes over, are demonstrated by gray areas.}
  \label{fig:UAV}
\end{figure*}

\subsubsection{Setup}
To demonstrate that the proposed algorithm, represented by closed-form expressions, can be implemented for real-time onboard computation, we conduct hardware-in-the-loop (HIL) experiments with the fixed-wing UAV. The hardware-in-the-loop (HIL) system is illustrated in Figure~\ref{fig:Semi-Physical}. In this architecture, the onboard computer manages all the computations of the motion planning algorithm, the autopilot functions as the flight control unit and generates commands for actuators based on our planned control inputs, and the simulator replicates the real-world flight environment while providing configuration feedback.

We consider a fixed-wing UAV with minimum turning radius $\rho=\SI{30}{m}$ and linear velocity bounded by $v_{\text{min}}=\SI{16}{m/s}$ and $v_{\text{max}}=\SI{18}{m/s}$. Due to the cruising speed constraint, this UAV has positive lower bound on its linear velocity, and hence can only converge to the target positive limit set rather than the target configuration. The parameters of the CVF are specified by $r_1=6\rho$, $r_2=12\rho$ and $r_3=18\rho$.

The HIL experiments are conducted following the procedure outlined below. Initially, the fixed-wing UAV takes off and maneuvers to the initial configuration $\bm{g}_0$, at which point the motion planning algorithm is activated, directing the UAV to the target configuration $\bm{g}_d$. Once the UAV reaches $\bm{g}_d$, the motion planning algorithm is deactivated while the autopilot guides the UAV to the subsequent initial configuration without landing. This process is repeated for all the examples in Table~\ref{tab:UAV}, where all the target configurations are encompassed by a common target positive limit set $\mathcal{L}_d^+=\{\bm{g} \mid r=r_2,\ \theta=\theta_r\}$.

\begin{table}[!hbtp]
  \centering
  \caption{Initial and Target Configurations for Fixed-Wing UAV}
  \begin{tabular}{| c | c c c | c c c |}
  \hline
    \textbf{Examples} & \multicolumn{3}{c|}{$\bm{g}_0=(x_0,y_0,\theta_0)$} & \multicolumn{3}{c|}{$\bm{g}_d=(x_d,y_d,\theta_d)$} \\ \hline
    \textbf{Exp.1} & (2.809 & 10.65 & -1.699)  & (-180.0 & -311.7 & -0.524) \\  \hline
    \textbf{Exp.2} & (-4.683 & -9.393 & 1.368) & (360.0 & 0 & 1.571) \\ \hline
    \textbf{Exp.3} & (10.43 & 0.146 & -2.830)  & (-180.0 & 311.7 & -2.618) \\ \hline
    \textbf{Exp.4} & (152.7 & 260.3 & 0.248)  & (-180.0 & -311.7 & -0.524) \\ \hline
    \textbf{Exp.5} & (-294.4 &  -6.373 &  2.273)  & (360.0 & 0 & 1.571) \\ \hline
    \textbf{Exp.6} & (149.0 & -262.9 &  -1.831)  & (-180.0 & 311.7 & -2.618) \\ \hline
    \textbf{Exp.7} & (338.3 & 520.6 &  -0.538)  & (-180.0 & -311.7 & -0.524) \\ \hline
    \textbf{Exp.8} & (-619.7 & 31.18 &   1.596)  & (360.0 & 0 & 1.571) \\ \hline
    \textbf{Exp.9} & (280.6 & -552.7 &  -2.673)  & (-180.0 & 311.7 & -2.618) \\ \hline
  \end{tabular}
  \label{tab:UAV}
\end{table}

\subsubsection{Setpoint Derivation}
The control inputs of linear velocity and angular velocity in 2D provided by our algorithm cannot be directly applied to 3D fixed-wing UAVs. Therefore, we first convert the linear and angular velocities into the corresponding setpoints of attitude and thrust for the autopilot platform as follows.

For simplicity, we assume that the attack angle is sufficiently small and that there is no slide-slip angle. Therefore, the body-fixed frame and the velocity frame of the fixed-wing UAV are equivalent.
The setpoint for the UAV's attitude is represented by the yaw-pitch-roll Euler angles, denoted by ($\alpha$, $\beta$, $\gamma$). The attitude matrix $\bm{R}\in \mathrm{SO}(3)$ is given by
\[
\bm{R}=\bm{R}_z(\alpha)\bm{R}_y(\beta)\bm{R}_x(\gamma),
\]
where $\bm{R}_i(\cdot)$ is the rotation matrix about the $i$-axis.

To deploy the proposed 2D motion planning algorithm in the experiment with UAV, we confine the flight in the horizontal plane at the height $h_d$ by adjusting the pitch angle $\beta$ towards the setpoint specified by
\begin{equation} \label{pitch_d}
    \beta_d = \frac{K_\beta (h-h_d)}{V},
\end{equation}
where $K_\beta$ is the gain, $h$ is the current flight height and $V$ is the airspeed.
When the experiment begins, the UAV takes off and then circles in the horizontal plane where $h=h_d$, waiting for the motion planning algorithm to take over. In this context, we assume that the pitch angle $\beta$ is already stabilized at zero in the subsequent derivation.

Next, we align the heading direction of the UAV with the CVF by setpoint of the yaw angle
\begin{equation} \label{yaw_d}
    \alpha_d = \arctan\frac{\left< \mathbf{T}\left([p_x,p_y]^\intercal\right) , \bm{e}_y^i \right>}{\left<\mathbf{T}\left([p_x,p_y]^\intercal\right) , \bm{e}_x^i\right>},
\end{equation}
where $\mathbf{T}(\cdot)$ is given by \eqref{CVF}, $p_x$ and $p_y$ are the first two components in the position vector of the UAV and $\bm{e}_x^i$ and $\bm{e}_y^i$ are the first two basis vectors of the inertial frame.

Subsequently, we aim to track the desired angular velocity with the time derivative of the yaw angle.
In accordance with the coordinated turn constraint defined by
\[ \dot{\alpha} = -\frac{a_{g}}{V}\tan{\gamma}\cos{\beta}\overset{\mathrm{\beta=0}}{=}-\frac{a_g}{V}\tan{\gamma}, \]
we derive the setpoint of roll angle that enables $\dot{\alpha}$ to follow $\omega$ as
\begin{equation} \label{roll_d}
    \gamma_d = -\arctan\frac{\omega V}{a_g},
\end{equation}
where $a_g$ is the gravitational acceleration and $\omega$ is determined by \eqref{omega}.
Finally, to track the planned linear velocity $v_x$ given by \eqref{vx} with airspeed $V$, we assign the thrust by
\begin{equation}
    F_T=\max\{(-K_T(V-v_x)+\dot{v}_x)m+F_D,0\},
\end{equation}
where $\dot{v}_x = \frac{\partial v_x}{\partial \bm{p}}^\intercal \dot{\bm{p}} + \frac{\partial v_x}{\partial \theta_e} \dot{\theta}_e$ according to \eqref{vx}
and $F_D$ is the drag.

\subsubsection{Results}
To evaluate the convergence towards the target positive limit set and the tracking performance of 3D setpoints, we define the distance error between the current position and the target positive limit set as $d_e= r_\delta-r_2 $ and present the attitude error with $\| \log_{\mathrm{SO}(3)}{\bm{R}_e} \|$ where $\bm{R}_e=\bm{R}\bm{R}_r^{-1}$ and $\bm{R}_r=\bm{R}_z(\alpha_d)\bm{R}_y(\beta_d)\bm{R}_z(\gamma_d)$.
The results of the HIL experiments on fixed-wing UAV are presented in Figures~\ref{fig:UAV trajectories}~and~\ref{fig:UAV}.

Despite the errors introduced by the simplification in the derivation of setpoints and dynamic uncertainties, the control inputs can be effectively tracked by the fixed-wing UAV. This is because the proposed algorithm addresses both the curvature constraint and the nonholonomic constraint, which are the most critical kinematic constraints for the UAV.
Consequently, both the distance error $d_e$ and the attitude error $\| \log_{\mathrm{SO}(3)}{\bm{R}_e} \|$ in Figure~\ref{fig:UAV} are stabilized as soon as the motion planning algorithm is activated. This confirms the convergence of the UAV to the target positive limit set and its arrival at the target configuration under curvature constraint, as shown in Figure~\ref{fig:UAV trajectories}.
Moreover, the results also demonstrate that the proposed algorithm can be implemented onboard in real-time, and can handle time-varying tasks where the target configuration switches during the flight, which is another advantage of the closed-form motion planning method.

\subsection{Parameter Selection for Different Robots} \label{subsec:parameter selection}
In this subsection, we explain the effect of the curvature constraint in the construction of the CVF and provide a practical strategy for radii selection in the CVF on different robot platforms.

The theoretical foundation for this parameter selection strategy is established in Theorems~\ref{thm F} and \ref{thm stabilization condition}, which provide a range of feasible values for the region radii, rather than designating specific values. This flexibility allows the region radii to be conveniently adjusted according to real-world scenarios, promising a wider applicability of the proposed method. For robots with better maneuverability and small turning radius (e.g., ground vehicles, unicycle-type robots), smaller and tighter region radii can be employed to create more compact CVFs with narrower transition regions. This enables faster convergence to the target positive limit set while maintaining desired curvature with a prescribed bound. In contrast, for robots with stricter dynamic constraints and larger turning radius (e.g., fixed-wing UAVs, large marine vessels), their CVFs are more sensitive to the change of control inputs. Therefore, larger and more disparate region radii are preferred to improve the robot's tracking performance with smoother control inputs.

\begin{figure}[htbp]
  \centering
  \subfigure[CVF for Ackermann UGV with minimum turning radius $\rho = \SI{0.6}{m}$]{\includegraphics[width=0.48\linewidth]{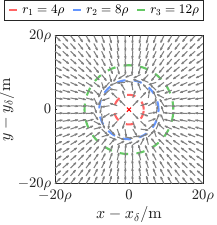} \label{Fig:CVF_UGV}}
  \hfill
  \subfigure[CVF for fixed-wing UAV with minimum turning radius $\rho = \SI{30}{m}$]{\includegraphics[width=0.48\linewidth]{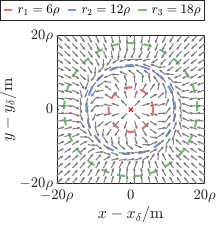} \label{Fig:CVF_UAV}}
  \caption{Comparison of the CVF for the different robots in hardware experiments, where the singular point $\bm{p}_\delta=\left[x_\delta,y_\delta\right]^\intercal$ is presented by the red cross.}
  \label{fig:experiment_comparison}
\end{figure}

The experimental validation in this section demonstrates the above principle. 
In Section~\ref{subsec:UGV}, the CVF for the Ackermann UGV with $\rho = \SI{0.6}{m}$ is constructed with compact parameters $r_1=4\rho=\SI{2.4}{m}$, $r_2=8\rho=\SI{4.8}{m}$, $r_3=12\rho=\SI{7.2}{m}$. 
It can be seen that the CVF evolves from source and sink to vortex in a narrower region with width $r_3-r_1=8\rho=\SI{4.8}{m}$ as shown in Figure~\ref{Fig:CVF_UGV}. 
In Section~\ref{subsec:HIL}, the fixed-wing UAV with $\rho = \SI{30}{m}$ requires more conservative parameters specified by $r_1=8\rho=\SI{240}{m}$, $r_2=12\rho=\SI{360}{m}$, $r_3=18\rho=\SI{540}{m}$ to accommodate its sensitivity to control input variations under strict curvature constraints. 
The CVF with larger region radii and wider transition region is shown in Figure~\ref{Fig:CVF_UAV}. 
The above parameter selection strategy ensures that the CVF-based motion planning is both theoretically sound and practically implementable across diverse robotic platforms.

\section{Conclusions}\label{sec:conclusion}

In this paper, we have investigated the curvature-constrained motion planning problem for nonholonomic robots. A novel framework that co-develops the curvature-constrained vector field and the saturated control laws with dynamic gain has been presented. The proposed methodology ensures that the robot's orientation aligns with the reference orientation and the robot's trajectory converges to the target positive limit set with bounded curvature almost globally.
The effectiveness of the obtained results is verified by both numerical simulations and hardware experiments on the Ackermann UGV and semi-physical fixed-wing UAV.
Potential future research includes the 3D CVF planning, and multi-robot  motion planning under both curvature and nonholonomic constraints.

\appendices
\section{Proof of Lemma~\ref{lemma_curvature}} \label{appendix1}
  The curvature of a curve can be computed by
  \begin{equation}
      \kappa=\frac{\|\dot{\boldsymbol{p}}\times \ddot{\boldsymbol{p}}\|}{\|\dot{\boldsymbol{p}}\|^3},
      \label{kappa 3d}
  \end{equation}
as provided in \cite[Chapter 13.3]{Stewart2011Calculus}, where the cross product in $\mathbb{R}^2$ is defined as  $[a,b]\times[c,d]\coloneq ad-bc$.

In the polar coordinates, the integral curve satisfies the following ordinary differential equation:
  \begin{equation}
    \mathbf{F} = \left[ \begin{matrix} F_r\\ F_\varphi   \end{matrix}\right]=\left[ \begin{matrix} \dot{r}\\ r\dot{\varphi}   \end{matrix}\right] 
    .
  \end{equation}
Moreover, the integral curve equation in Cartesian coordinates can be transformed from the polar coordinates into
\begin{equation}\label{polare}
	\begin{aligned}
		\bm{p} &= \left[ \begin{matrix} r\cos{\varphi}\\r\sin{\varphi}\end{matrix}\right]
		= R(\varphi)\left[ \begin{matrix} r\\0\end{matrix}\right],\\
		\dot{\bm{p}} &= \begin{bmatrix}F_x\\F_y\end{bmatrix} = R(\varphi) \begin{bmatrix}F_r\\F_\varphi\end{bmatrix},
	\end{aligned}
\end{equation}
where the 2-D rotation matrix $R(\varphi)$ is defined as
\begin{equation}
	R(\varphi) = \left[ \begin{matrix} \cos{\varphi} & -\sin{\varphi}\\ \sin{\varphi} & \cos{\varphi} \end{matrix}\right].
\end{equation}
In light of the above equations, we further derive the second-order derivative of $\bm{p}$ as follows:
\begin{equation} \label{sub equations}
\begin{aligned}
    \ddot{\bm{p}} &= \frac{\mathrm{d}}{\mathrm{d}t}(R(\varphi))\begin{bmatrix} F_r\\ F_\varphi \end{bmatrix} + R(\varphi) \frac{\mathrm{d}}{\mathrm{d}t} \begin{bmatrix} F_r\\ F_\varphi \end{bmatrix} \\
    &= R(\varphi)\begin{bmatrix} 0 & -\dot{\varphi}\\ \dot{\varphi} & 0 \end{bmatrix}\begin{bmatrix} F_r\\ F_\varphi \end{bmatrix} + R(\varphi)\begin{bmatrix} \dot{F}_r\\ \dot{F}_\varphi \end{bmatrix} \\
    &= R(\varphi) \begin{bmatrix}
        0 & -\frac{F_\varphi}{r}\\
        \frac{F_\varphi}{r} & 0 
    \end{bmatrix} 
    \begin{bmatrix} F_r\\ F_\varphi \end{bmatrix} \\
    &\quad + R(\varphi) \begin{bmatrix}
        \frac{\partial F_r}{\partial r} & \frac{\partial F_r}{r\partial\varphi} \\
        \frac{\partial F_\varphi}{\partial r} & \frac{\partial F_\varphi}{r\partial\varphi}
    \end{bmatrix}  
    \begin{bmatrix} \dot{r}\\ r\dot{\varphi} \end{bmatrix} \\
    &= R(\varphi)\begin{bmatrix}
        \frac{\partial F_r}{\partial r} & \frac{\partial F_r}{r\partial\varphi} - \frac{F_\varphi}{r}\\
        \frac{\partial F_\varphi}{\partial r} + \frac{F_\varphi}{r} & \frac{\partial F_\varphi}{r\partial\varphi}
    \end{bmatrix}\mathbf{F}.
\end{aligned}
\end{equation}
By substituting \ref{polare} and \ref{sub equations} into \eqref{kappa 3d}, we can obtain \eqref{VF_k}.

\section{Proof of Theorem~\ref{thm F}} \label{appendix_F}
  The proof for Theorem~\ref{thm F} is divided into three parts, where we firstly demonstrate the smoothness of the blended VF $\mathbf{F}$, then show the existence of stable limit cycle and finally derive the condition for the integral curves to have a prescribed bound on the curvature.

\begin{figure}[h]
    \centering
    \includegraphics[width=0.9\linewidth]{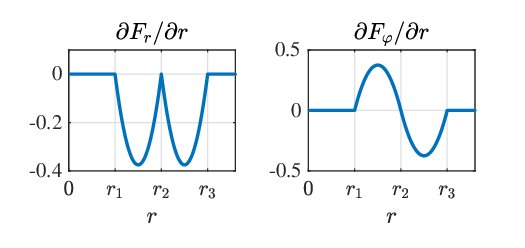}
    \caption{Plot of $\partial F_r/\partial \varphi$ and $\partial F_\varphi / \partial \varphi$.}
    \label{fig:partial derivatives}
  \end{figure}

  1) According to \eqref{out,v} and \eqref{v,in}, we know $\mathbf{F}\ne\bm{0}$, $\forall \bm{p}\in \mathcal{D}$ since both $\mathbf{F}_\text{out}$ and $\mathbf{F}_\text{in}$ are orthogonal to $\mathbf{F}_\text{v}$. Therefore, $\bm{p}=\bm{0}$ is the unique singular point for $\mathbf{F}$. Subsequently, the VF $\mathbf{F}$ is shown to be of class $C^1$ by computing the first order derivative of the components of $\mathbf{F}$, where continuous components are given by
  \begin{subequations}
    \begin{align}
      F_r=\left\{\begin{matrix}
        1, & 0\le r < r_1\\
        \lambda(s_2), & r_1 \le r < r_2\\
        \lambda(s_3)-1, & r_2 \le r < r_3\\
        -1, & r_3 \le r < \infty
      \end{matrix}\right., \label{F_r} \\
      F_\varphi=\left\{\begin{matrix}
        0, & 0\le r < r_1\\
        1-\lambda(s_2), & r_1 \le r < r_2\\
        \lambda(s_3), & r_2 \le r < r_3\\
        0, & r_3 \le r < \infty
      \end{matrix}\right..
    \end{align}
  \end{subequations}
  Since the VF is symmetric about $\bm{p}=\bm{0}$, we know $\partial F_r/\partial \varphi=\partial F_\varphi / \partial \varphi =0$. The derivatives of $F_r$ and $F_\varphi$ w.r.t. $r$ are given by 
  \[
  \begin{aligned}
    \frac{\partial F_r}{\partial r}=\left\{\begin{matrix}
      0, & 0\le r < r_1\\
      (6s_2^2-6s_2)/(r_2-r_1), & r_1 \le r < r_2\\
      (6s_3^2-6s_3)/(r_3-r_2), & r_2 \le r < r_3\\
      0, & r_3 \le r < \infty
    \end{matrix}\right.,\\
    \frac{\partial F_\varphi}{\partial r}=\left\{\begin{matrix}
      0, & 0\le r < r_1\\
      -(6s_2^2-6s_2)/(r_2-r_1), & r_1 \le r < r_2\\
      (6s_3^2-6s_3)/(r_3-r_2), & r_2 \le r < r_3\\
      0, & r_3 \le r < \infty
    \end{matrix}\right., 
  \end{aligned}
  \]
  which are continuous as shown in Figure~\ref{fig:partial derivatives}. Therefore, the VF $\mathbf{F}$ is of class $C^1$ over $\mathcal{D}$.

  2) To prove that $\mathcal{O}_2$ is an almost stable limit cycle for integral curves with kinematics in Definition~\ref{def_IC}, we firstly show the existence of the limit cycle and then analyze its stability.
  Consider a closed annular region $\mathcal{A}_{inv}=\{\bm{p}\mid \|\bm{p}\|\in[\delta_1,\ \delta_2],\ 0<\delta_1<r_2,\ \delta_2>r_2 \}$ containing $\mathcal{O}_2$ and denote the inner and outer boundaries of $\mathcal{A}_{inv}$ as $\mathcal{I}_{inv}$ and $\mathcal{O}_{inv}$, respectively. Since both boundaries are smooth, the tangent cone of $\mathcal{A}_{inv}$ in $\bm{p}\in \partial \mathcal{A}_{inv}$ is the tangent halfspace of $\partial \mathcal{A}_{inv}$ depicted by
  \begin{equation}
  \begin{aligned}
    \mathcal{C}_{\mathcal{A}_{inv}}(\bm{p})=\{\bm{\xi}\mid \left<\bm{p},\bm{\xi}\right>\ge0\},\ \forall \bm{p}\in \mathcal{I}_{inv},\\
    \mathcal{C}_{\mathcal{A}_{inv}}(\bm{p})=\{\bm{\xi}\mid \left<\bm{p},\bm{\xi}\right>\le0\},\ \forall \bm{p}\in \mathcal{O}_{inv}.
  \end{aligned}
  \end{equation}
  Given \eqref{the_VF}, it can be verified that $\mathbf{F}(\bm{p})\in\mathcal{C}_{\mathcal{A}_{inv}}(\bm{p})$, $\forall \bm{p}\in \partial \mathcal{A}_{inv}$, implying $\mathcal{A}_{inv}$ is positively invariant according to Nagumo's theorem \cite{nagumo1942lage}.
  Additionally, the positively invariant annular region $\mathcal{A}_{inv}$ contains at least one limit cycle since it contains no singular point and $\left<\mathbf{F}(\bm{p}),\bm{p}\right>\neq0$, $\forall \bm{p}\in\partial \mathcal{A}_{inv}$ indicates the existence of forward orbit from the boundary \cite[Theorem 1.179]{chicone2006ordinary}.

  Next, we show $\mathcal{O}_2$ is the unique and stable limit cycle on $\mathcal{D}$. Define $r_d\coloneq \|\bm{\zeta}\|-r_2 \in [0,\infty)$, whose time derivative is given by
  \begin{equation}\label{d_rd}
    \dot{r}_d=\mathrm{d}\|\bm{\zeta}\|/\mathrm{dt}=F_r.
  \end{equation}
  According to \eqref{F_r}, there hold $\dot{r}_d>0$ for $-r_2<r_d<0$ and $\dot{r}_d<0$ for  $r_d>0$.
  Subsequently, we choose a positive definite Lyapunov function candidate:
  \begin{equation}
    V(r_d)=\frac{1}{2}r_d^2,
  \end{equation}
  and the time derivative is given by
  \begin{equation}
    \dot{V}(r_d)=r_d\dot{r}_d<0,
  \end{equation}
  implying $\lim_{t\to\infty}\|\bm{\zeta}\|=r_2$ for any initial condition $\bm{\zeta}_0\neq\bm{0}$ and thereby $\mathcal{O}_2$ is almost globally stable.

  3) Since the VF $\mathbf{F}$ is of class $C^1$, we can conclude that the curvature is continuous over $\mathcal{D}$ according to Lemma~\ref{lemma_curvature}. To show the boundedness of the curvature, we first discuss the trivial case where $\bm{p}\in\mathcal{A}_1\setminus \bm{0}$. In this case, there holds that $F_r=1$ and $F_\varphi=0$ and $\bm{K}_F=\bm{0}$. Substituting these values into \eqref{VF_k}, we have $\kappa_\mathbf{F}(\bm{p})=0$, $\forall \bm{p}\in\mathcal{A}_1\setminus \bm{0}$. Similarly we can obtain the same result that $\kappa_\mathbf{F}(\bm{p})=0$, $\forall \bm{p}\in\mathcal{A}_4$, where $F_r=-1$ and $F_\varphi=0$.

  Next, consider $\bm{p}\in\mathcal{A}_2$, where $r_1 \le r < r_2$, $F_r = \lambda(s_2)$, $F_\varphi = \tilde{\lambda}(s_2)$, $\lambda(s_2)=2s_2^3-3s_2^2+1$ and $s_2 = (r-r_1)(r_2-r_1)\in [0,1)$.
  Substituting the above equations into \eqref{VF_k} yields
  \begin{equation}
    \begin{aligned}\label{k_A2}
      &\kappa_\mathbf{F}(\bm{p})\\
      & = \left| \frac{1-\lambda(s_2)}{(\lambda(s_2)^2+\tilde{\lambda}(s_2)^2)^{1/2}}\frac{1}{r} + \frac{\frac{\partial \lambda(s_2)}{\partial s_2}\lambda(s_2)}{(\lambda(s_2)^2+\tilde{\lambda}(s_2)^2)^{3/2}}\frac{1}{r_2-r_1} \right| \\
      & = \lvert \frac{k_1(s_2)}{r} + \frac{k_2(s_2)}{r_2-r_1} \rvert
        \stackrel{r \ge r_1}{\le} \lvert \frac{k_1(s_2)}{r_1} \rvert + \lvert \frac{k_2(s_2)}{r_2-r_1} \rvert\\
      &\stackrel{\eqref{curvature condition 1}}{\le} \frac{\lvert k_1(s_2) \rvert + \lvert k_2(s_2) \rvert}{r_2-r_1}\stackrel{\eqref{curvature condition 2}}{\le} \frac{\lvert k_1(s_2) \rvert + \lvert k_2(s_2) \rvert}{3\rho} \le \bar{\kappa},
    \end{aligned}
  \end{equation}
  where $k_1$, $k_2$ are single variable functions of $s_2$ and the term $\lvert k_1(s_2) \rvert + \lvert k_2(s_2) \rvert$ is bounded by 3 on $s_2\in[0,1)$, as illustrated in Figure~\ref{fig:k1,k2}. Therefore, the curvature of the integral curves in $\mathcal{A}_2$ is bounded by $\bar{\kappa}$.
  Finally, we consider $\bm{p}\in\mathcal{A}_3$, where $r\ge r_2$, $F_r=-\tilde{\lambda}(s_3)$, $F_\varphi=\lambda(s_3)$, $\lambda(s_3)=2s_3^3-3s_3^2+1$ and $s_3=(r-r_2)/(r_3-r_2)$.
  In this case, the curvature can be written as
  \begin{equation}
  \begin{aligned}\label{k_A3}
    & \kappa_\mathbf{F}(\bm{p})\\
    & = \lvert \frac{\lambda(s_3)}{(\lambda(s_3)^2+\tilde{\lambda}(s_3)^2)^{1/2}}\frac{1}{r} + \frac{\frac{\partial \lambda(s_3)}{\partial s_3}(1-\lambda(s_3))}{(\lambda(s_3)^2+\tilde{\lambda}(s_3)^2)^{3/2}}\frac{1}{r_3-r_2} \rvert \\
    & = \lvert \frac{k_3(s_3)}{r} + \frac{k_4(s_3)}{r_3-r_2} \rvert
    \stackrel{r \ge r_2}{\le} \lvert \frac{k_3(s_3)}{r_2} \rvert + \lvert \frac{k_4(s_3)}{r_3-r_2} \rvert\\
    &\stackrel{\eqref{curvature condition 1}}{\le} \frac{\lvert k_3(s_3) \rvert + \lvert k_4(s_3) \rvert}{r_3-r_2}\stackrel{\eqref{curvature condition 2}}{\le} \frac{\lvert k_3(s_3) \rvert + \lvert k_4(s_3) \rvert}{3\rho} \le \bar{\kappa},
  \end{aligned}
  \end{equation}
  where $\lvert k_3(s_3) \rvert + \lvert k_4(s_3) \rvert$ is bounded by 3 on $s_3\in[0,1)$.
  By analogous inequality scaling, we can show that $\kappa_\mathbf{F}(\bm{p})\le\bar{\kappa}$ still holds for $\bm{p}\in\mathcal{A}_3$.
  \begin{figure}[h]
    \centering
    \includegraphics[width=0.9\linewidth]{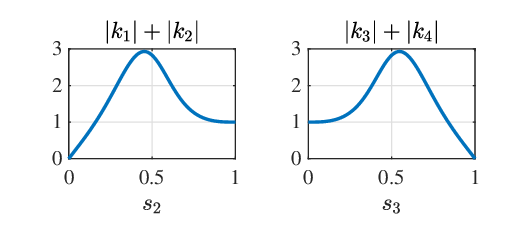}
    \caption{Plot of $\lvert k_1(s_2) \rvert + \lvert k_2(s_2)\rvert$ on $s_2 \in [0,1)$ and $\lvert k_3(s_3) \rvert + \lvert k_4(s_3)\rvert$ on $s_3 \in [0,1)$.}
    \label{fig:k1,k2}
  \end{figure}

\section{Proof of Theorem~\ref{thm T}} \label{appendix_T}
  In this proof, we show that the proposed CVF has a bounded and continuous curvature and the target position $\bm{p}_d$ belongs to the positive limit set of integral curves and $\mathbf{T}(\bm{p}_d)/\|\mathbf{T}(\bm{p}_d)\|=[\cos{\theta_d},\sin{\theta_d}]^\intercal$.

  1) Upper bounded continuous curvature.

  According to \cite[Chapter 1.5]{chicone2006ordinary}, the normalization operation reparameterizes the time of the integral curves of $\mathbf{T}$ from $\mathbf{F}$. 
  In other words, the shape of integral curves of $\mathbf{T}$ with $\bm{\zeta}_0=\bm{p}$ remains the same as those of $\mathbf{F}$ with $\bm{\zeta}_0=\bm{p}+\bm{p}_\delta$. Therefore, the integral curves from $\mathbf{F}$ to $\mathbf{T}$ are translated by $\bm{p}_\delta$.
  In addition, the translation is a positive isometry, which preserves the curvature of the integral curves \cite{banchoff2016Chap3.4}. Therefore, the integral curves of $\mathbf{T}$ still have continuous curvature bounded by $\bar{\kappa}$ as shown in Theorem~\ref{thm F}.

  2) Positive limit set of integral curves of $\mathbf{T}$.

  In above discussion, we have shown that the integral curves of $\mathbf{T}$ are translated from those of $\mathbf{F}$ by $\bm{p}_\delta$. Consequently, the stable limit cycle for integral curves of $\mathbf{T}$, which is a special case of the positive limit set, is depicted by $\mathcal{L}^+_\mathbf{T}(\bm{\zeta}_0)=\{\bm{p} \mid \|\bm{p}-\bm{p}_\delta\|=r_2\}$, $\bm{\zeta}_0\neq \bm{p}_\delta$. 
  Since $\bm{p}_\delta=\bm{p}_d-\bm{p}_{\theta_d}$ and $\|\bm{p}_{\theta_d}\|=r_2$, we can conclude that $\bm{p}_d\in\mathcal{L}^+_\mathbf{T}(\bm{\zeta}_0)$, $\forall \bm{\zeta}_0\neq \bm{p}_\delta$.

  3) $\mathbf{T}(\bm{p}_d)$ aligning with $\theta_d$.

  Given \eqref{CVF}, we transform the components of $\mathbf{F}$ from $(r,\varphi)$ coordinates into Cartesian coordinates by rotation matrix $R(\varphi)$ and there holds that
  \begin{equation}
  \begin{aligned}
    &\lim_{\bm{p}\to\bm{p}_d}\frac{\mathbf{T}(\bm{p})}{\|\mathbf{T}(\bm{p})\|}
    =\lim_{\bm{q}\to-\bm{p}_{\theta_d}}\frac{\mathbf{F}(\bm{q})}{\|\mathbf{F}(\bm{q})\|}\\
    &=\lim_{\substack{r\to r_2\\ \varphi\to\theta_d-\frac{\pi}{2}}} R(\varphi)\left[\begin{matrix}
    F_r(r)\\F_\varphi(r)
    \end{matrix}\right]/\|R(\varphi)\left[\begin{matrix}
    F_r(r)\\F_\varphi(r)
    \end{matrix}\right]\|\\
    &=\left[\begin{matrix}
    \cos{(\theta_d-\frac{\pi}{2})} &-\sin{(\theta_d-\frac{\pi}{2})}\\
    \sin{(\theta_d-\frac{\pi}{2})} &\cos{(\theta_d-\frac{\pi}{2})}
    \end{matrix}\right]
    \left[\begin{matrix}
    0\\1
    \end{matrix}\right]=\left[\begin{matrix}
    \cos{\theta_d}\\\sin{\theta_d}
    \end{matrix}\right].
  \end{aligned}
  \end{equation}

\section{Kinematics of Reference Orientation} \label{appendix2}
Knowing $\|\mathbf{T}\|=1$ and $\theta_r=\arctan{T_y/T_x}$, we have
\begin{equation}\label{omega_r initial}
  \omega_r=T_x\dot{T}_y-T_y\dot{T}_x=\left[\begin{matrix}
    -T_y &T_x\end{matrix}\right]\left[\begin{matrix}
    \dot{T}_x\\ \dot{T}_y
    \end{matrix}\right].
\end{equation}
In this section, we consider the polar coordinate centered given by 
\[
\begin{aligned}
    r_\delta &= \|\bm{p}-\bm{p}_\delta\|, \\
    \varphi_\delta &= \arctan{\frac{\left<\bm{p}-\bm{p}_\delta,\bm{e}_y\right>}{\left<\bm{p}-\bm{p}_\delta,\bm{e}_x\right>}},
\end{aligned}
\]
which is centered at the singular point $\bm{p}_\delta$ with basis $\left\{\bm{e}_{r_\delta},\bm{e}_{\varphi_\delta}\right\}$.
The transformation of $\mathbf{T}$ between Cartesian coordinate $(x,y)$ and polar coordinate $(r_\delta,\varphi_\delta)$ is given by
\[
  \left[\begin{matrix}T_x\\ T_y\end{matrix}\right]
  = R(\varphi_\delta)\left[\begin{matrix}T_{r_\delta}\\ T_{\varphi_\delta}\end{matrix}\right],
\]
where the components of $\mathbf{T}$ in polar coordinates are defined as $T_{r_\delta} = \left<\mathbf{T},\bm{e}_{r_\delta} \right>$ and $T_{\varphi_\delta} = \left<\mathbf{T},\bm{e}_{\varphi_\delta} \right>$.
Correspondingly, we have
\[
  \left[\begin{matrix}
  -T_y &T_x\end{matrix}\right]
  = \left[\begin{matrix}
  T_{r_\delta} &T_{\varphi_\delta}\end{matrix}\right]R{\left(-(\varphi_\delta+\frac{\pi}{2})\right)}.
\]
Next we compute $\dot{\mathbf{T}}$ as follows:
\[
  \begin{aligned}
  &\dot{\mathbf{T}} = \left[\begin{matrix}
  \dot{T}_x\\ \dot{T}_y
  \end{matrix}\right]
  =\frac{\mathrm{d}}{\mathrm{d}t} \left(R{(\varphi_\delta)}\left[\begin{matrix}T_{r_\delta}\\ T_{\varphi_\delta}\end{matrix}\right] \right)\\
  & =\frac{\mathrm{d}R{(\varphi_\delta)}}{\mathrm{d}t}
  \left[\begin{matrix}T_{r_\delta}\\ T_{\varphi_\delta}\end{matrix}\right]
  + R{(\varphi_\delta)}
  \frac{\mathrm{d}}{\mathrm{d}t} \left[\begin{matrix}T_{r_\delta}\\ T_{\varphi_\delta} \end{matrix}\right].
  \end{aligned}
\]
Substituting the above equations into \eqref{omega_r initial}, we have
\[
  \omega_r = \dot{\varphi}_\delta + (T_{r_\delta}\dot{T}_{\varphi_\delta}-T_{\varphi_\delta}\dot{T}_{r_\delta}),
\]
where the first term is introduced by the rotation of the polar coordinate $(r_\delta,\varphi_\delta)$ while the second term is due to the robot's relative motion in the polar coordinate. Furthermore, by decomposing the linear velocity $\bm{v}_x$ into $\bm{v}_x=v_x^\parallel \bm{e}_{r_\delta} + v_x^\perp \bm{e}_{\varphi_\delta}$, the above equation can be rewritten as
\begin{equation}\label{omega_r}
\begin{aligned}
    \omega_r
    &= \frac{1}{r_\delta}\frac{\partial \theta_r}{\partial \varphi_\delta}v_x^\perp + \frac{\partial \theta_r}{\partial r_\delta}v_x^\parallel\\
    &= \frac{1}{r_\delta}v_x^\perp + \theta_{r,r_\delta}v_x^\parallel\\
    &= A(r_\delta)v_x\cos{\Delta \theta},
\end{aligned}
\end{equation}
where $\Delta \theta = \theta - \theta_\nabla$, $\theta_\nabla=\arctan{\frac{\left< \nabla \theta_r, \bm{e}_y \right>}{\left< \nabla \theta_r, \bm{e}_x \right>}}$ and $A(r_\delta)=\|\nabla\theta_r\|$ denote the direction and magnitude of the gradient $\nabla\theta_r$, respectively.
Moreover, given the partial derivative of $\theta_r$ w.r.t. $r_\delta$:
\begin{equation}\label{g(r)}
\theta_{r,r_\delta}(r_\delta) = \frac{\partial \theta_r}{\partial r_\delta} = \left\{\begin{matrix}
  &0 &\bm{p}-\bm{p}_\delta\in \mathcal{A}_1,\\
  &\frac{6s_2-6s_2^2}{(r_2-r_1)(2\lambda^2_2-2\lambda_2+1)} &\bm{p}-\bm{p}_\delta\in \mathcal{A}_2,\\
  &\frac{6s_3-6s_3^2}{(r_3-r_2)(2\lambda^2_3-2\lambda_3+1)} &\bm{p}-\bm{p}_\delta\in \mathcal{A}_3,\\
  &0 &\bm{p}-\bm{p}_\delta\in \mathcal{A}_4,
\end{matrix}\right.
\end{equation}
where $s_i=(r-r_{i-1})/(r_{i}-r_{i-1})$, $\lambda_i=2s^3_i-3s^2_i+1$ for $i=2,3$ and $\mathcal{A}_i$ are defined in \eqref{decomposition}, the direction and magnitude of $\nabla\theta_r$ can be computed by $\theta_\nabla=\varphi_\delta+\mathrm{arccot}(r_\delta\theta_{r,r_\delta})$ and $A(r_\delta)=\sqrt{1/r_\delta^2+\theta_{r,r_\delta}^2}$, respectively.

\section{Proof of Theorem~\ref{thm dynamic gain}}\label{appendix_S}
  When the angular velocity gain is specified by \eqref{dynamic gain}, the bound of $\omega_0$ in \eqref{proportional} can be computed by
  \begin{equation}
  \begin{aligned}\label{w0 bound}
        \lvert\omega_0\rvert
        & \le k_\omega\lvert\theta_e\rvert+ \lvert \omega_r \rvert\\
        & \le v_x(\bar{\kappa}-k(r_\delta)\lvert\cos{\Delta \theta}\rvert) + A(r_\delta) v_x \lvert \cos{\Delta \theta} \rvert\\
        & = \bar{\omega} + \left(A(r_\delta)-k(r_\delta)\right)v_x\lvert \cos{\Delta \theta} \rvert\\
        & = \bar{\omega}_0.
  \end{aligned}
  \end{equation}
  Furthermore, the saturation region on the configuration space SE(2) can be estimated by
  \begin{equation} \label{reduced inequality}
        \begin{aligned}
        \mathcal{S} &= \left\{ \bm{g} \mid r_\delta < \rho \right\} 
        = \left\{\bm{g} \mid k(r_\delta)<A(r_\delta) \right\}\\
        &\supseteq \left\{\bm{g} \mid \bar{\omega}_0 > \bar{\omega} \right\}
        \supseteq \left\{\bm{g} \mid \left| \omega_0 \right| >\bar{\omega}\right\}, 
        \end{aligned}
  \end{equation}
  since there is $k(r_\delta)\le \bar{\kappa}<A(r_\delta)=1/r_\delta$, $\forall r_\delta < \rho$ according to \eqref{k<kappa} and \eqref{k}. Therefore, the angular velocity can only possibly saturate when the robot is in the open circular region with radius $\rho$, centered at the singular point $\bm{p}_\delta$.

\section{Proof of Theorem~\ref{thm stabilization condition}}\label{appendix_theta_e}

  In this proof, we firstly explain the composition of the non-converging set.
  The subsets $\mathcal{N}_{\mathbf{T}}$ and $\mathcal{N}_\theta$ are non-converging because the orientation error $\theta_e=\theta-\theta_r$ is not well-defined for $\bm{g}_0 \in \mathcal{N}_{\mathbf{T}} \cup \mathcal{N}_\theta$. As for $\bm{g}\in\mathcal{N}_{k_\omega}$, it can be shown that $k_\omega=0$. In this case, there hold that $\omega = \omega_r = \bar{\omega}$ and $\lvert \theta_e \rvert = \pi/2$, meaning that the system will stay on one of the following limit cycle: $\mathcal{C}_\infty^+=\left\{\bm{g} \mid r_\delta=\rho,\ \theta_e =\pi/2\right\}$ or $\mathcal{C}_\infty^-=\left\{\bm{g} \mid r_\delta=\rho,\ \theta_e = -\pi/2\right\}$.
  Therefore, the set $\mathcal{N}_{k_\omega}$ is also non-converging.
  However, these limit cycles are unstable since any arbitrary perturbation on the configuration will induce $k_{\omega}>0$, driving the system away from the limit cycles.
  As a result, we can obtain that $k_\omega>0$, $\forall t\ge 0$ when $\bm{g}_0 \notin \mathcal{N}_{k_\omega}$. Since each subset of $\mathcal{N}$ is of measure zero, we can conclude that the non-converging set $\mathcal{N}$ is also of measure zero from the sub-additivity property of measures.

  Moreover, to ensure the existence of the nonlinear function $k(\cdot)$ satisfying \eqref{k<kappa} and \eqref{k}, the CVF should satisfy  
  \begin{equation}\label{existence condition}
    A(r_\delta)\le\bar{\kappa},\ \forall r \ge \rho.
  \end{equation}
  For $\rho \le r_\delta < r_1$ or $r_\delta \ge r_3$, where $\theta_{r,r_\delta}=0$ and hence $A(r_\delta)=1/r_\delta$ according to \eqref{g(r)}, the condition \eqref{existence condition} is satisfied.
  Otherwise, for $r_{i-1} \le r_\delta < r_{i}$, $i=2,3$, where $\theta_{r,r_\delta}$ is non-zero, we show that \eqref{existence condition} still holds by the following inequality:
  \[\begin{aligned}
    A(r_\delta)
    &=\sqrt{\frac{1}{r_\delta^2}+\theta^2_{r,r_\delta}} \le \frac{1}{r_\delta}+\theta_{r,r_\delta}\\
    &\le \max_{r_{i-1} \le r_\delta < r_{i}}{\frac{1}{r_\delta}}+\max_{r_{i-1} \le r_\delta < r_{i}}{\theta_{r,r_\delta}}\\
    &\le \frac{1}{r_{i-1}} + \frac{1}{r_{i}-r_{i-1}} \stackrel{\eqref{stabilization condition}}{\le} \bar{\kappa},
  \end{aligned}\]
  where $\max_{r_{i-1} \le r_\delta < r_{i}}{\theta_{r,r_\delta}} \le \frac{1}{r_{i}-r_{i-1}}$ can be examined by simple algebra. 

  When there exists a dynamic gain that ensures the angular velocity only saturates in $\mathcal{S}$, we consider the Lyapunov candidate $V_\theta=\frac{1}{2}\theta_e^2$ and discuss its evolution twofold.
  On the one hand, when the angular velocity control law \eqref{omega} does not saturate, there are $\omega=\omega_0$ and $\dot{\theta}_e = \omega - \omega_r = -k_\omega \theta_e$. Therefore, the time derivative of $V_\theta$ is given by $\dot{V}_\theta = -k_\omega \theta_e^2 < 0$.
  On the other hand, if the angular velocity saturates, implying that $\bm{g}(t) \in \mathcal{S}$, we know $\theta_{grad}=\theta_r+\pi/2$ according to \eqref{CVF} and \eqref{omega_r}. Then, the kinematics of $\theta_r$ can be rewritten into
  \[
    \omega_r = \dot{\theta}_r = A(r_\delta)v_x\sin{\theta_e},
  \]
  which means the feedback term $-k_\omega\theta_e$ and the feedforward term $\omega_r$ in \eqref{proportional} have different signs. In addition, the dynamic gain \eqref{dynamic gain} ensures that $\lvert k_\omega \theta_e \rvert \le \bar{\omega}$. Therefore, we can conclude that \eqref{omega} saturates, i.e. $\lvert\omega_0\rvert = \lvert-k_\omega\theta_e + \omega_r\rvert > \bar{\omega}$, if and only if $\lvert \omega_r \rvert > \bar{\omega} + \lvert k_\omega \theta_e \rvert$.
  When $\theta_e > 0$ and the angular velocity saturates, there holds that
  \[
    \left.\begin{matrix}
      \omega_r > \bar{\omega} + k_\omega \theta_e\\
      \omega = \bar{\omega}
    \end{matrix}\right\}
    \Rightarrow
    \dot{\theta}_e = \omega - \omega_r < -k_\omega \theta_e < 0,
  \]
  and thereby $\dot{V}_\theta = \theta_e\dot{\theta}_e < -k_\omega \theta_e^2 < 0$. By similar computations, we can obtain that $\dot{V}_\theta < 0$ for $\theta_e < 0$.
  Therefore, $\dot{V}_\theta$ is always negative definite, no matter whether the angular velocity \eqref{omega} saturates or not, which implies that the orientation error $\theta_e$ is monotonically stabilized.

\section{Proof of Theorem~\ref{thm convergence}}\label{appendix_convergence}
\begin{figure}
      \centering
      \includegraphics[width=0.9\linewidth]{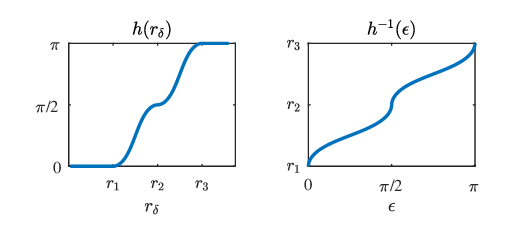}
      \caption{Plot of $h(\cdot)$ and $h^{-1}(\cdot)$.}\label{fig:h}
  \end{figure}

  Since the orientation error $\theta_e$ is proved to be monotonically stabilized to zero, we can claim that there exists a finite time $T$ such that $\theta_e(t) \le \pi/2$, $\forall t \ge T$. Without loss of generality, assume that the initial condition $\bm{g}_0 \in \mathcal{N}$ has initial orientation error $\theta_e \le \pi/2$. In this case, the kinematics of the closed-loop system can be reformulated into the following cascade system:
  \begin{subequations}\label{cascade}
    \begin{align}
      &\dot{r}_\delta=f_1(r_\delta,\theta_e)=v_x\cos{(\theta_r-\varphi_\delta+\theta_e)}, \label{dot_r} \\
      &\dot{\theta}_e = f_2(\theta_e)=\omega - \omega_r. \label{dot_theta}
    \end{align}
  \end{subequations}
  For the sake of simplicity, we denote $h(r_\delta)=\theta_r-\varphi_\delta$ and the plot of $h(r_\delta)$ w.r.t. $r_\delta>0$ is presented in Figure~\ref{fig:h}.
  Now we regard $\theta_e$ in \eqref{dot_theta} as the input of \eqref{dot_r}, and define the Lyapunov candidate $V_\delta=(r_\delta-r_2)^2/2$ for \eqref{dot_r}, whose derivative is
    \begin{equation}\label{d_V_delta}
    \dot{V}_\delta=v_x(r_\delta-r_2)\cos{(h(r_\delta)+\theta_e)}.
    \end{equation}
    Analyzing $h(r_\delta)$ and \eqref{d_V_delta}, we have
    \[\dot{V}_\delta<0,\ \forall \lvert r_\delta -r_2 \rvert \ge
    \gamma(\lvert \theta_e \rvert), \]
    where $\gamma(\lvert \theta_e \rvert)=\max{\left\{\gamma_1,\gamma_2\right\}}$, $\gamma_1(\lvert\theta_e\rvert)=r_2-h^{-1}(\pi/2-\lvert\theta_e\rvert))\le r_2-r_1$ and $\gamma_2(\lvert\theta_e\rvert)=h^{-1}(\pi/2+\lvert\theta_e\rvert))-r_2\le r_3-r_2$ are class $\mathcal{K}$ functions as illustrated in Figure~\ref{fig:h}.
  Referring to \cite[Theorem 4.19]{khalil2002nonlinear}, the above Lyapunov analysis suggests that the subsystem \eqref{dot_r} with bounded input $\lvert\theta_e\rvert\le\pi/2$ is input-to-state stable (ISS).
  Furthermore, since Theorem~\ref{thm stabilization condition} indicates that the origin of \eqref{dot_theta} is asymptotically stable, we conclude the cascade system \eqref{cascade} converges to the equilibrium where $r_\delta = r_2$, $\theta_e=0$ \cite[Lemma 4.7]{khalil2002nonlinear}.
  
  The ISS of the cascade system \eqref{cascade} can be interpreted twofold. 
  For the robot with $v_{min}>0$, the closed-loop system tracks the CVF and moves along the integral curve of CVF. According to Theorem~\ref{thm T}, the integral curves of CVF converge to limit cycle that contains the target position $\bm{p}_d$ and the CVF aligns with the target orientation $\theta_d$ at $\bm{p}_d$. Therefore, the closed-loop system is guaranteed to converge to the target positive limit set $\mathcal{L}_d^+=\left\{\bm{g} \mid r_\delta =r_2,\ \theta_e=0 \right\}\ni\bm{g}_d$ and the planning objective \eqref{planning obj} is satisfied.
  As for the robot with $v_{min}=0$, its linear velocity $v_x$ vanishes as the robot approaches the target configuration $\bm{g}_d$ along the integral curve of CVF. In this case, the target positive limit set is $\mathcal{L}_d^+=\left\{ \bm{g}_d \right\}$ and the objective \eqref{planning obj} is simplified as $\lim_{t\to\infty}\bm{g}(t)=\bm{g}_d$.

\section*{Acknowledgment}
The authors would like to thank Xiwang Dong, Haoze Dong, and Tianqing Liu for their assistance with the implementation of the experiments on the Ackermann UGV and the fixed-wing UAV.

\bibliographystyle{IEEEtran}
\bibliography{mybibfile}

@book{lee2012introduction,
    address = {New York},
    author = {Lee, John M.},
    title = {Introduction to Smooth Manifolds},
    publisher = {Springer},
    year = {2012},
    edition = {2nd},
    doi = {10.1007/978-1-4419-9982-5},
    isbn = {978-1-4419-9982-5}
}

@book{White2016Fluid,
  author    = {F. M. White},
  title     = {Fluid Mechanics},
  edition   = {8th},
  publisher = {McGraw-Hill},
  address   = {New York, NY, USA},
  year      = {2016},
}

@article{yao2023guiding,
  author={Yao, Weijia and de Marina, Héctor García and Sun, Zhiyong and Cao, Ming},
  journal={IEEE Transactions on Robotics}, 
  title={Guiding Vector Fields for the Distributed Motion Coordination of Mobile Robots}, 
  year={2023},
  volume={39},
  number={2},
  pages={1119-1135},
  publisher={IEEE}
}

@article{pathak2005integrated,
  title={An integrated path-planning and control approach for nonholonomic unicycles using switched local potentials},
  author={Pathak, Kaustubh and Agrawal, Sunil Kumar},
  journal={IEEE Transactions on Robotics},
  volume={21},
  number={6},
  pages={1201--1208},
  year={2005},
  publisher={IEEE}
}

@inproceedings{masoud2009harmonic,
  title={A harmonic potential field approach for navigating a rigid, nonholonomic robot in a cluttered environment},
  author={Masoud, Ahmad A},
  booktitle={Proceedings of 2009 IEEE International Conference on Robotics and Automation},
  pages={3993--3999},
  year={2009}
}

@inproceedings{lindemann2007smooth,
  title={Smooth feedback for car-like vehicles in polygonal environments},
  author={Lindemann, Stephen R and LaValle, Steven M},
  booktitle={Proceedings of 2007 IEEE International Conference on Robotics and Automation},
  pages={3104--3109},
  year={2007}
}

@inproceedings{lindemann2006real,
  title={Real time feedback control for nonholonomic mobile robots with obstacles},
  author={Lindemann, Stephen R and Hussein, Islam I and LaValle, Steven M},
  booktitle={Proceedings of 45th IEEE Conference on Decision and Control},
  pages={2406--2411},
  year={2006}
}

@inproceedings{balluchi1996optimal,
  title={Optimal feedback control of {D}ubins' car tracking circular reference paths},
  author={Balluchi, Andrea and Sou{\'e}res, Philippe},
  booktitle={Proceedings of 35th IEEE Conference on Decision and Control},
  volume={3},
  pages={3558--3563},
  year={1996}
}

@inproceedings{scheuer1997continuous,
  title={Continuous-curvature path planning for car-like vehicles},
  author={Scheuer, Alexis and Fraichard, Th},
  booktitle={Proceedings of 1997 IEEE/RSJ International Conference on Intelligent Robot and Systems},
  volume={2},
  pages={997--1003},
  year={1997}
}

@inproceedings{qiao2024motion,
  author={Qiao, Yike and He, Xiaodong and Li, Zhongkui},
  booktitle={2024 IEEE 63rd Conference on Decision and Control}, 
  title={Motion Planning of 3{D} Nonholonomic Robots via Curvature-Constrained Vector Fields}, 
  year={2024},
  pages={5807-5812},
  doi={10.1109/CDC56724.2024.10886648}
}

@article{pothen2017curvature,
  title={Curvature-constrained {L}yapunov vector field for standoff target tracking},
  author={Pothen, Abin Alex and Ratnoo, Ashwini},
  journal={Journal of Guidance, Control, and Dynamics},
  volume={40},
  number={10},
  pages={2729--2736},
  year={2017},
  publisher={American Institute of Aeronautics and Astronautics}
}

@inproceedings{shivam2021curvature,
  title={Curvature-Constrained Vector Field for Path Following Guidance},
  author={Shivam, Amit and Ratnoo, Ashwini},
  booktitle={Proceedings of 2021 International Conference on Unmanned Aircraft Systems},
  pages={853--857},
  year={2021}
}

@article{he2025novel,
  title={A novel vector-field-based motion planning algorithm for 3{D} nonholonomic robots},
  author={He, Xiaodong and Yao, Weijia and Sun, Zhiyong and Li, Zhongkui},
  journal={Automatica},
  volume={172},
  pages={111996},
  year={2025},
  publisher={Elsevier}
}

@article{chen2020dubins,
  title={On {D}ubins paths to a circle},
  author={Chen, Zheng},
  journal={Automatica},
  volume={117},
  pages={108996},
  year={2020},
  publisher={Elsevier}
}

@inproceedings{xu2021autonomous,
  title={Autonomous vehicle motion planning via recurrent spline optimization},
  author={Xu, Wenda and Wang, Qian and Dolan, John M},
  booktitle={Proceedings of 2021 IEEE International Conference on Robotics and Automation},
  pages={7730--7736},
  year={2021}
}

@article{song2018t,
   author={Song, Junnan and Gupta, Shalabh and Wettergren, Thomas A.},
  journal={IEEE Robotics and Automation Letters}, 
  title={T$^\star$: Time-Optimal Risk-Aware Motion Planning for Curvature-Constrained Vehicles}, 
  year={2019},
  volume={4},
  number={1},
  pages={33-40},
}

@inproceedings{liu2024low,
  title={Low-cost differential flatness identification for trajectory planning and tracking of small fixed-wing {UAV}s in dense environments},
  author={Liu, Tianqing and Wang, Mengyun and Niu, Yifeng and Li, Jie and Zhou, Han},
  booktitle={Proceedings of 2024 International Conference on Unmanned Aircraft Systems},
  pages={905--910},
  year={2024}
}

@article{yao2021singularity,
  title={Singularity-free guiding vector field for robot navigation},
  author={Yao, Weijia and de Marina, H{\'e}ctor Garcia and Lin, Bohuan and Cao, Ming},
  journal={IEEE Transactions on Robotics},
  volume={37},
  number={4},
  pages={1206--1221},
  year={2021},
  publisher={IEEE}
}

@inproceedings{panagou2010dipole,
  title={Dipole-like fields for stabilization of systems with Pfaffian constraints},
  author={Panagou, Dimitra and Tanner, Herbert G and Kyriakopoulos, Kostas J},
  booktitle={Proceedings of 2010 IEEE International Conference on Robotics and Automation},
  pages={4499--4504},
  year={2010}
}

@inproceedings{duan2014planning,
  title={Planning locally optimal, curvature-constrained trajectories in 3{D} using sequential convex optimization},
  author={Duan, Yan and Patil, Sachin and Schulman, John and Goldberg, Ken and Abbeel, Pieter},
  booktitle={Proceedings of 2014 IEEE International Conference on Robotics and Automation},
  pages={5889--5895},
  year={2014}
}

@inproceedings{panagou2014motion,
  title={Motion planning and collision avoidance using navigation vector fields},
  author={Panagou, Dimitra},
  booktitle={Proceedings of 2014 IEEE International Conference on Robotics and Automation},
  pages={2513--2518},
  year={2014}
}

@article{egeland1996feedback,
  title={Feedback control of a nonholonomic underwater vehicle with a constant desired configuration},
  author={Egeland, Olav and Dalsmo, Morten and Soerdalen, Ole Jakob},
  journal={The International Journal of Robotics Research},
  volume={15},
  number={1},
  pages={24--35},
  year={1996},
  publisher={Sage Publications Sage CA: Thousand Oaks, CA}
}

@article{aiello2022fixed,
  title={Fixed-wing {UAV} energy efficient 3{D} path planning in cluttered environments},
  author={Aiello, Giuseppe and Valavanis, Kimon P and Rizzo, Alessandro},
  journal={Journal of Intelligent \& Robotic Systems},
  volume={105},
  number={3},
  pages={60},
  year={2022},
  publisher={Springer}
}

@article{levin2019real,
  title={Real-time motion planning with a fixed-wing {UAV} using an agile maneuver space},
  author={Levin, Joshua M and Nahon, Meyer and Paranjape, Aditya A},
  journal={Autonomous Robots},
  volume={43},
  number={8},
  pages={2111--2130},
  year={2019},
  publisher={Springer}
}

@article{du2016robust,
  title={Robust dynamic positioning of ships with disturbances under input saturation},
  author={Du, Jialu and Hu, Xin and Krsti{\'c}, Miroslav and Sun, Yuqing},
  journal={Automatica},
  volume={73},
  pages={207--214},
  year={2016},
  publisher={Elsevier}
}

@article{li2019robust,
  title={Robust time-varying formation control for underactuated autonomous underwater vehicles with disturbances under input saturation},
  author={Li, Jian and Du, Jialu and Chang, Wen-Jer},
  journal={Ocean Engineering},
  volume={179},
  pages={180--188},
  year={2019},
  publisher={Elsevier}
}

@article{shima2002time,
  title={Time-varying linear pursuit-evasion game models with bounded controls},
  author={Shima, Tal and Shinar, Josef},
  journal={Journal of Guidance, Control, and Dynamics},
  volume={25},
  number={3},
  pages={425--432},
  year={2002}
}

@article{taub2013intercept,
  title={Intercept angle missile guidance under time varying acceleration bounds},
  author={Taub, Ilan and Shima, Tal},
  journal={Journal of Guidance, Control, and Dynamics},
  volume={36},
  number={3},
  pages={686--699},
  year={2013},
  publisher={American Institute of Aeronautics and Astronautics}
}

@article{chen2019shortest,
  title={Shortest {D}ubins paths through three points},
  author={Chen, Zheng and Shima, Tal},
  journal={Automatica},
  volume={105},
  pages={368--375},
  year={2019},
  publisher={Elsevier}
}

@article{chen2023elongation,
  title={Elongation of curvature-bounded path},
  author={Chen, Zheng and Wang, Kun and Shi, Heng},
  journal={Automatica},
  volume={151},
  pages={110936},
  year={2023},
  publisher={Elsevier}
}

@article{tanner2003nonholonomic,
  title={Nonholonomic navigation and control of cooperating mobile manipulators},
  author={Tanner, Herbert G and Loizou, Savvas G and Kyriakopoulos, Kostas J},
  journal={IEEE Transactions on Robotics and Automation},
  volume={19},
  number={1},
  pages={53--64},
  year={2003},
  publisher={IEEE}
}

@ARTICLE{he2024simultaneous,
   author={He, Xiaodong and Li, Zhongkui},
  journal={IEEE Transactions on Automatic Control}, 
  title={Simultaneous Position and Orientation Planning of Nonholonomic Multirobot Systems: A Dynamic Vector Field Approach}, 
  year={2024},
  volume={69},
  number={12},
  pages={8354-8369},
  doi={10.1109/TAC.2024.3406475}
}

@book{khalil2002nonlinear,
  title={Nonlinear Systems},
  author={Khalil, H.K.},
  isbn={9780130673893},
  lccn={95045804},
  series={Pearson Education},
  year={2002},
  publisher={Prentice Hall}
}

@article{nagumo1942lage,
  title={{\"U}ber die lage der integralkurven gew{\"o}hnlicher differentialgleichungen},
  author={Nagumo, Mitio},
  journal={Proceedings of the Physico-Mathematical Society of Japan. 3rd Series},
  volume={24},
  pages={551--559},
  year={1942},
  publisher={THE PHYSICAL SOCIETY OF JAPAN, The Mathematical Society of Japan}
}

@article{li2013finite,
  title={Finite-time consensus and collision avoidance control algorithms for multiple {AUV}s},
  author={Li, Shihua and Wang, Xiangyu},
  journal={Automatica},
  volume={49},
  number={11},
  pages={3359--3367},
  year={2013},
  publisher={Elsevier}
}

@article{he2024roto,
  title={Roto-translation invariant formation of fixed-wing {UAV}s in 3{D}: Feasibility and control},
  author={He, Xiaodong and Li, Zhongkui and Wang, Xiangke and Geng, Zhiyong},
  journal={Automatica},
  volume={161},
  pages={111492},
  year={2024},
  publisher={Elsevier}
}

@article{karaman2011sampling,
  title={Sampling-based algorithms for optimal motion planning},
  author={Karaman, Sertac and Frazzoli, Emilio},
  journal={The International Journal of Robotics Research},
  volume={30},
  number={7},
  pages={846--894},
  year={2011},
  publisher={Sage Publications}
}

@article{murray1993nonholonomic,
  title={Nonholonomic motion planning: Steering using sinusoids},
  author={Murray, Richard M and Sastry, S Shankar},
  journal={IEEE Transactions on Automatic Control},
  volume={38},
  number={5},
  pages={700--716},
  year={1993},
  publisher={IEEE}
}

@article{Hussein2008Optimal,
  title={Optimal control of underactuated nonholonomic mechanical systems},
  author={Hussein, Islam I and Bloch, Anthony M},
  journal={IEEE Transactions on Automatic Control},
  volume={53},
  number={3},
  pages={668--682},
  year={2008},
  publisher={IEEE}
}

@article{lindemann2009simple,
  title={Simple and efficient algorithms for computing smooth, collision-free feedback laws over given cell decompositions},
  author={Lindemann, Stephen R and LaValle, Steven M},
  journal={The International Journal of Robotics Research},
  volume={28},
  number={5},
  pages={600--621},
  year={2009},
  publisher={SAGE Publications}
}

@ARTICLE{Panagou2017Distributed,
  author={Panagou, Dimitra},
  journal={IEEE Transactions on Automatic Control}, 
  title={A Distributed Feedback Motion Planning Protocol for Multiple Unicycle Agents of Different Classes}, 
  year={2017},
  volume={62},
  number={3},
  pages={1178-1193},
  }

@article{dubins1957curves,
  title={On curves of minimal length with a constraint on average curvature, and with prescribed initial and terminal positions and tangents},
  author={Dubins, Lester E},
  journal={American Journal of mathematics},
  volume={79},
  number={3},
  pages={497--516},
  year={1957},
  publisher={JSTOR}
}

@article{lau2015fluid,
  title={Fluid motion planner for nonholonomic 3-{D} mobile robots with kinematic constraints},
  author={Lau, Darwin and Eden, Jonathan and Oetomo, Denny},
  journal={IEEE Transactions on Robotics},
  volume={31},
  number={6},
  pages={1537--1547},
  year={2015},
  publisher={IEEE}
}

@Book{Stewart2011Calculus,
  Title                    = {Calculus},
  Author                   = {J.~Stewart},
  Publisher                = {Cengage Learning},
  Year                     = {2011},
  Address = {Boston, MA, USA}
}

@incollection{banchoff2016Chap3.4,
  title = {Curves in space: Local properties},
  chapter = {3.4},
  booktitle={Differential geometry of curves and surfaces},
  author={Banchoff, Thomas F and Lovett, Stephen},
  year={2016},
  publisher={CRC Press},
  address       = {Boca Raton, FL, USA},
  edition = {2},
  pages = {97--101}
}

@Book{chicone2006ordinary,
  Title                    = {Ordinary differential equations with applications},
  Author                   = {Chicone, Carmen},
  Publisher                = {Springer Science \& Business Media},
  Year                     = {2006},
  Address                  = {New York, NY, USA},
  edition                  = {2}
}

\begin{IEEEbiography}[{\includegraphics[width=1in,height=1.25in,clip,keepaspectratio]{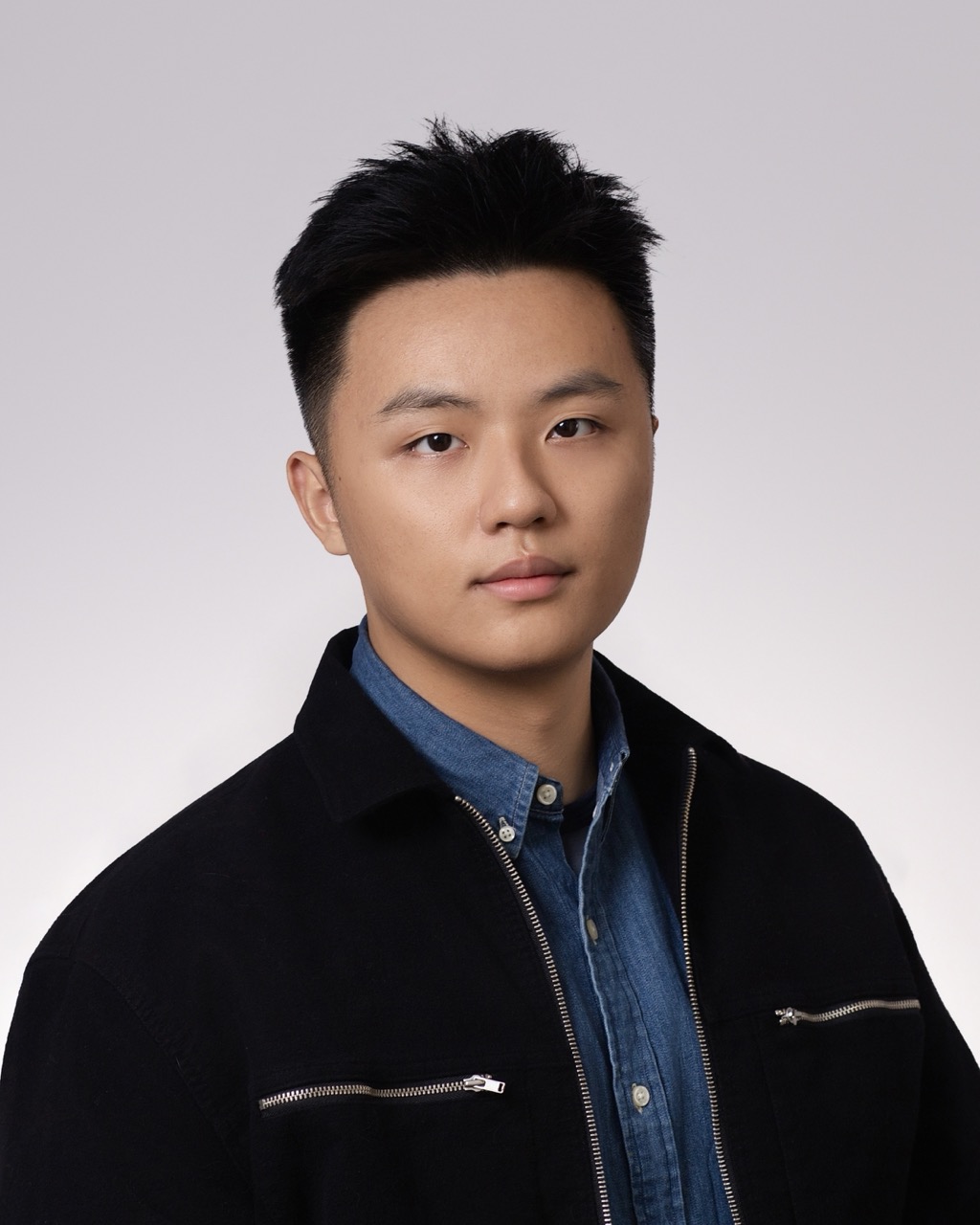}}]{Yike Qiao}
  (Student Member, IEEE) is working toward the Ph.D. degree in dynamical systems and control with the School of Advanced Manufacturing and Robotics, Peking University, Beijing, China.
  His research interests include motion planning and control of mobile robots under kinematic constraints.
\end{IEEEbiography}

\begin{IEEEbiography}[{\includegraphics[width=1in,height=1.25in,clip,keepaspectratio]{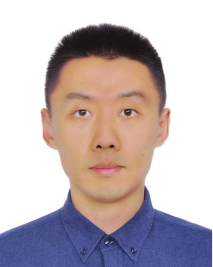}}]{Xiaodong He}
  (Member, IEEE) received the B.Eng. degree in mechanical engineering from China University of Petroleum, Qingdao, China, in 2016, and the Ph.D. degree in dynamics and control from the Peking University, Beijing, China, in 2021. From 2021 to 2024, he was a Postdoctoral Researcher and an Associate Researcher with the College of Engineering, Peking University. Since 2024, he has been an Associate Professor with the School of Automation and Electrical Engineering, University of Science and Technology Beijing, Beijing, China. His research interests include motion planning and control of nonholonomic systems.

  Dr. He was a finalist for the Zhang Si-Ying Outstanding Youth Paper Award of Chinese Control and Decision Conference (CCDC) in 2023.
\end{IEEEbiography}

\begin{IEEEbiography}[{\includegraphics[width=1in,height=1.25in,clip,keepaspectratio]{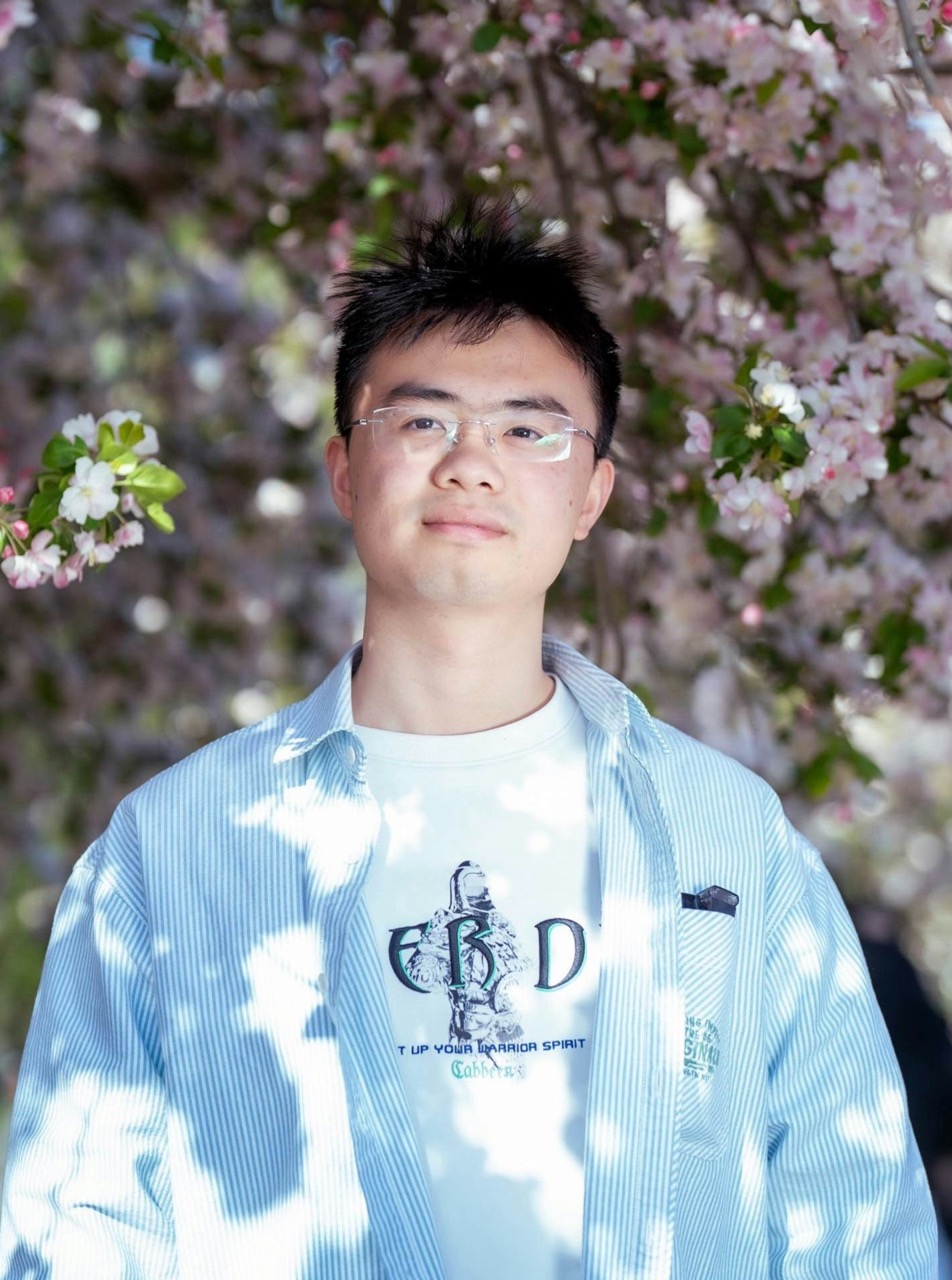}}]{An Zhuo} 
  is a Ph.D. candidate at the School of Advanced Manufacturing and Robotics, Peking University. He received his bachelor's degree from the National University of Defense Technology in 2024. His research interests include unmanned swarm systems, cooperative control, and decision making.
\end{IEEEbiography}

\begin{IEEEbiography}[{\includegraphics[width=1in,height=1.25in,clip,keepaspectratio]{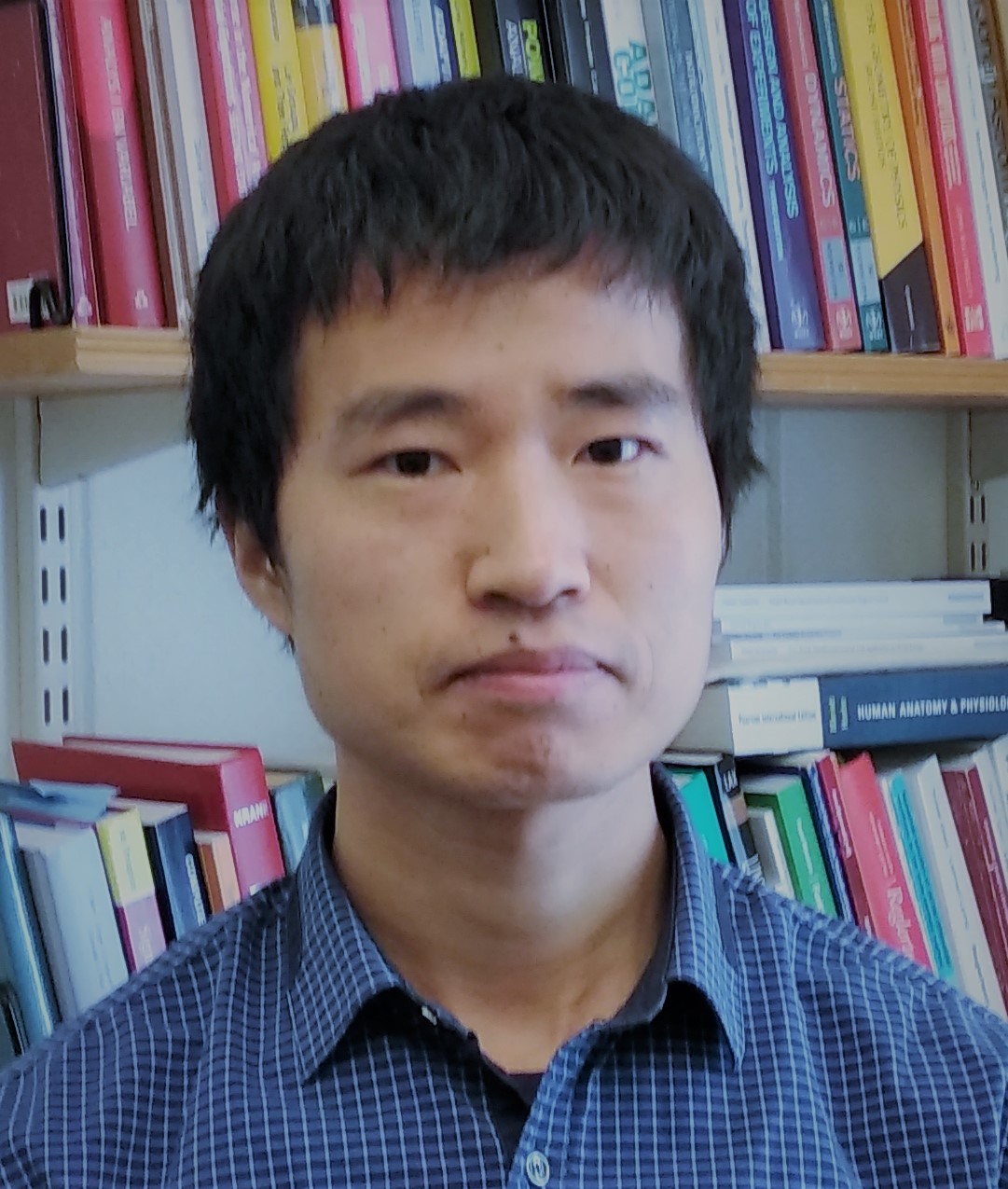}}]{Zhiyong Sun}
	(Member, IEEE) received the Ph.D. degree from The Australian National University (ANU), Canberra ACT, Australia, in February 2017. He was a Research Fellow/Lecturer with the Research School of Engineering, ANU, from 2017 to 2018. From June 2018 to January 2020, he worked as a postdoctoral researcher at the Department of Automatic Control, Lund University of Sweden. From January 2020 to July 2024, he worked with Eindhoven University of Technology (TU/e) as an assistant professor. He joined Peking University of China as a faculty member in the summer of 2024.
    
  His research interests include multi-agent systems, control of autonomous formations, distributed control and optimization. He has authored the book titled \textit{Cooperative Coordination and Formation Control for Multiagent Systems} (Springer, 2018). He has won the Springer Best Ph.D. Thesis Award, and several best paper and student paper awards from IEEE CDC, AuCC, ICRA, CCTA and ICCA.
\end{IEEEbiography}

\begin{IEEEbiography}[{\includegraphics[width=1in,height=1.25in,clip,keepaspectratio]{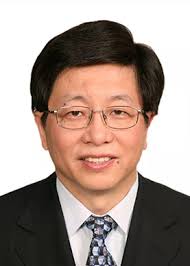}}]{Weimin Bao}
  received the B.E. degree in electronic engineering from Xidian University, Xi'an, China, in 1982. He is currently a CAS Member of China and a Professor with Xidian University. His research interests include space launch vehicle and control engineering.
\end{IEEEbiography}

\begin{IEEEbiography}[{\includegraphics[width=1in,height=1.25in,clip,keepaspectratio]{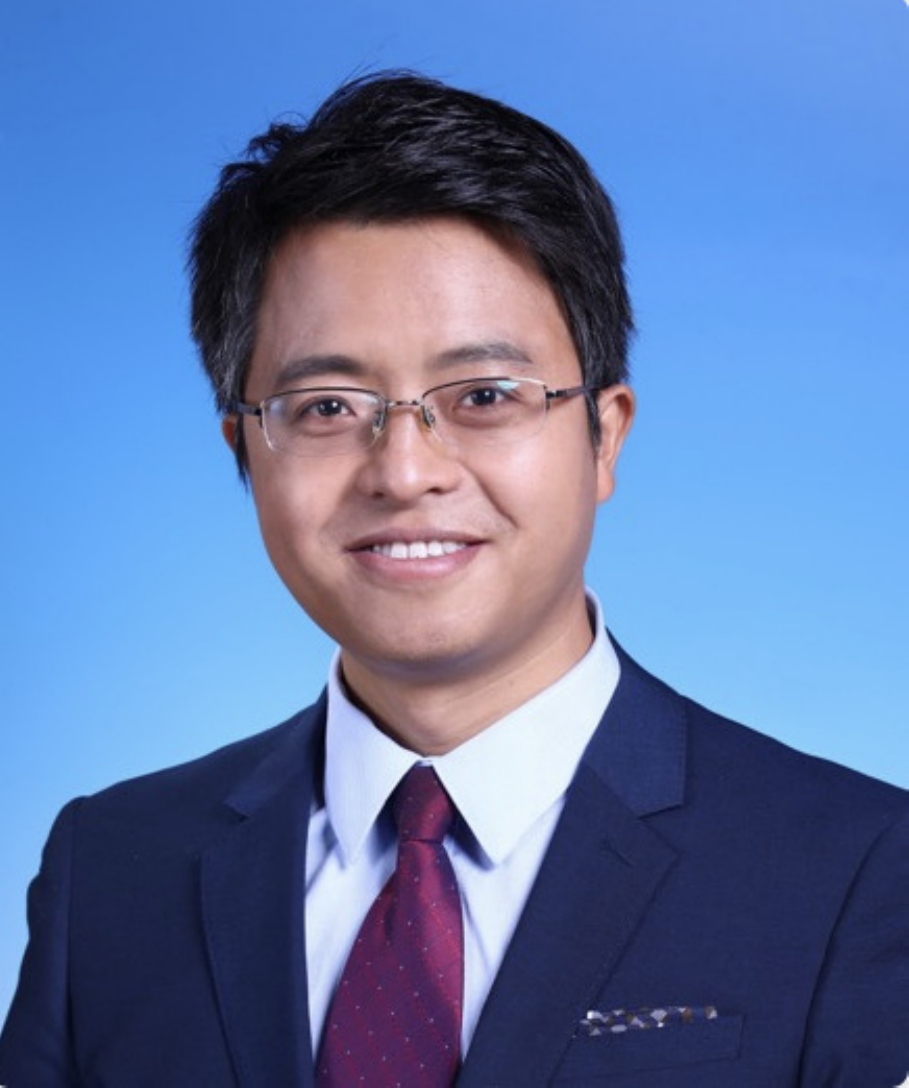}}]{Zhongkui Li}
  (Senior Member, IEEE) received the B.S. degree in space engineering from the National University of Defense Technology, China, in 2005, and his Ph.D. degree in dynamical systems and control from Peking University, China, in 2010. 
  
  Since 2013, Dr. Li has been with the Peking University, China, where he is currently a Full Professor with the School of Advanced Manufacturing and Robotics. His current research interests include cooperative control and planning of multi-agent systems.
  
  Dr. Li was the recipient of the China National Science Funds for Distinguished Young Scientists in 2024, the State Natural Science Award of China in 2015, the Natural Science Award of the Ministry of Education of China in 2022 and 2011, and the National Excellent Doctoral Thesis Award of China in 2012. His coauthored papers received the IET Control Theory \& Applications Premium Award in 2013 and the Best Paper Award of Journal of Systems Science \& Complexity in 2012. He serves/has served as an Associate Editor of IEEE Transactions on Automatic Control, and several other journals.
\end{IEEEbiography}

\end{document}